%% file: 3I.tex
\documentclass[]{aastex7}
\usepackage{ulem}
\usepackage[version=4]{mhchem}

\newcommand\subs[1]{\textsubscript{#1}}
\newcommand\sups[1]{\textsuperscript{#1}}
\newcommand\rh[1]{\textcolor{black}{{\textit{r}\subs{\textit{H}}}#1}}

\newcommand\Ju[1]{\textcolor{black}{{\textit{J}}#1}}

\newcommand\tkin[1]{\textcolor{black}{{\textit{T}\subs{kin}}#1}}

\newcommand\kms[1]{\textcolor{black}{{km\,s$^{-1}$}#1}}
\newcommand\ms[1]{\textcolor{black}{{m\,s$^{-1}$}#1}}
\newcommand\ps[1]{\textcolor{black}{{s$^{-1}$}#1}}

\definecolor{gold}{rgb}{0.64,0.54,0.29}

\received{2025 November 25}
\revised{2026 February 4}
\accepted{2026 February 9}
\submitjournal{The Astrophysical Journal Letters}

\shorttitle{CH$_3$OH and HCN in Interstellar Comet 3I/ATLAS}
\shortauthors{Roth et al.}
\begin{document}

%\title{Leveraging the ALMA Atacama Compact Array for Cometary Science: Mapping HCN, CH\subs{3}OH, H\subs{2}CO, CS, and HNC in Comet C/2015 ER61 (PanSTARRS)}

%\title{Methanol and Hydrogen Cyanide in Interstellar Object 3I/ATLAS Measured with the ALMA Atacama Compact Array}
\title{CH$_3$OH and HCN in Interstellar Comet 3I/ATLAS Mapped with the ALMA Atacama Compact Array: Distinct Outgassing Behaviors and a Remarkably High CH$_3$OH/HCN Production Rate Ratio}

\correspondingauthor{Nathan X. Roth}
\email{nathaniel.x.roth@nasa.gov}

\author[0000-0002-6006-9574]{Nathan X. Roth}
\affiliation{Solar System Exploration Division, Astrochemistry Laboratory Code 691, NASA Goddard Space Flight Center, 8800 Greenbelt Rd, Greenbelt, MD 20771, USA}
\affiliation{Department of Physics, American University, 4400 Massachusetts Ave NW, Washington, DC 20016, USA}
\email{nathaniel.x.roth@nasa.gov}

\author[0000-0001-8233-2436]{Martin A. Cordiner}
\affiliation{Solar System Exploration Division, Astrochemistry Laboratory Code 691, NASA Goddard Space Flight Center, 8800 Greenbelt Rd, Greenbelt, MD 20771, USA}
\affiliation{Department of Physics, The Catholic University of America, 620 Michigan Ave., N.E. Washington, DC 20064, USA}
\email{martin.cordiner@nasa.gov}

\author{Dominique Bockelée-Morvan}
\affiliation{LIRA, Observatoire de Paris, Université PSL, CNRS, Sorbonne Université, Université Paris Cité, 5 place Jules Janssen, 92195 Meudon, France}
\email{dominique.bockelee@obspm.fr}

\author[0000-0003-2414-5370]{Nicolas Biver}
\affiliation{LIRA, Observatoire de Paris, Université PSL, CNRS, Sorbonne Université, Université Paris Cité, 5 place Jules Janssen, 92195 Meudon, France}
\email{nicolas.biver@obspm.fr}

\author{Jacques Crovisier}
\affiliation{LIRA, Observatoire de Paris, Université PSL, CNRS, Sorbonne Université, Université Paris Cité, 5 place Jules Janssen, 92195 Meudon, France}
\email{jacques.crovisier@obspm.fr}

\author[0000-0001-7694-4129]{Stefanie N. Milam}
\affiliation{Solar System Exploration Division, Astrochemistry Laboratory Code 691, NASA Goddard Space Flight Center, 8800 Greenbelt Rd, Greenbelt, MD 20771, USA}
\email{stefanie.n.milam@nasa.gov}

\author[0000-0001-7168-1577]{Emmanuel Lellouch}
\affiliation{LIRA, Observatoire de Paris, Université PSL, CNRS, Sorbonne Université, Université Paris Cité, 5 place Jules Janssen, 92195 Meudon, France}
\email{jacques.crovisier@obspm.fr}

\author[0000-0002-1123-983X]{Pablo Santos-Sanz}
\affiliation{Instituto de Astrofísica de Andalucía (CSIC),
Glorieta de la Astronomía s/n, 18008-Granada, Spain}
\email{psantos@iaa.es}

\author[0000-0002-0500-4700]{Dariusz C. Lis}
\affiliation{Jet Propulsion Laboratory, California Institute of Technology, 4800 Oak Grove Drive, Pasadena, CA 91109, USA}
\email{dariusz.c.lis@jpl.nasa.gov}

\author[0000-0001-8642-1786]{Chunhua Qi}
\affiliation{Institute for Astrophysical Research, Boston University, 725 Commonwealth Avenue, Boston, MA 02215, USA}
\email{cqi1@bu.edu}

\author[0000-0001-6192-3181]{K. D. Foster}
\affiliation{Department of Chemistry, University of Virginia, Charlottesville, VA 22904, USA}
\email{kfoster@virginia.edu}

\author[0000-0002-1545-2136]{Jérémie Boissier}
\affiliation{Institut de Radioastronomie Millimetrique, 300 rue de la Piscine, F-38406
Saint Martin d'Heres, France}
\email{boissier@iram.fr}

\author[0000-0002-2026-8157]{Kenji Furuya}
\affiliation{RIKEN Pioneering Research Institute, 2-1 Hirosawa, Wako-shi, Saitama 351-0198, Japan}
\email{kenji.furuya@riken.jp}

\author{Raphael Moreno}
\affiliation{LIRA, Observatoire de Paris, Université PSL, CNRS, Sorbonne Université, Université Paris Cité, 5 place Jules Janssen, 92195 Meudon, France}
\email{raphael.moreno@obspm.fr}

\author[0000-0001-6752-5109]{Steven B. Charnley}
\affiliation{Solar System Exploration Division, Astrochemistry Laboratory Code 691, NASA Goddard Space Flight Center, 8800 Greenbelt Rd, Greenbelt, MD 20771, USA}
\email{steven.b.charnley@nasa.gov}

\author[0000-0001-9479-9287]{Anthony J. Remijan}
\affiliation{National Radio Astronomy Observatory, 520 Edgemont Rd, Charlottesville, VA 22903, USA}
\email{aremijan@nrao.edu}

\author[0000-0002-4336-0730]{Yi-Jehng Kuan}
\affiliation{Center of Astronomy and Gravitation, and Department of Earth Sciences, National Taiwan Normal University, 88 Section 4, Ting-Chou Road, Taipei 116, Taiwan}
\affiliation{Academia Sinica Institute of Astronomy and Astrophysics, P.O. Box 23-141, Taipei 106, Taiwan}
\email{kuan@ntnu.edu.tw}

\author[0009-0009-9925-7001]{Lillian X. Hart}
\affiliation{Department of Physics, University of Maryland, College Park, MD}
\email{lillianxh1@gmail.com}



%\nocollaboration{0}

\input{Abstract}



\input{body}

\begin{acknowledgments}
This work was supported by the Planetary Science Division Internal Scientist Funding Program through the Fundamental Laboratory Research (FLaRe) work package (N.X.R., S.N.M., M.A.C., S.B.C.), as well as the NASA Astrobiology Institute through the Goddard Center for Astrobiology (proposal 13-13NAI7-0032; S.N.M., M.A.C., S.B.C.). P.S-S. acknowledges financial support from the Spanish I+D+i project PID2022-139555NB-I00 (TNO-JWST) and from the Severo Ochoa grant CEX2021-001131-S funded by MICIU/AEI/10.13039/501100011033. Part of this research was carried out at the Jet Propulsion Laboratory, California Institute of Technology, under a contract with the National Aeronautics and Space Administration (80NM0018D0004). D.C.L. acknowledges financial support from the National Aeronautics and Space Administration (NASA) Astrophysics Data Analysis Program (ADAP). It makes use of the following ALMA data: ADS/JAO.ALMA \#2024.A.00049.S, \#2024.1.00477.T, and \#2024.1.00137.T. ALMA is a partnership of ESO (representing its member states), NSF (USA) and NINS (Japan), together with NRC (Canada), NSTC and ASIAA (Taiwan), and KASI (Republic of Korea), in cooperation with the Republic of Chile. The Joint ALMA Observatory is operated by ESO, AUI/NRAO and NAOJ. The National Radio Astronomy Observatory is a facility of the National Science Foundation operated under cooperative agreement by Associated Universities, Inc. We thank an anonymous referee for their feedback, which we feel improved the manuscript.
\end{acknowledgments}

\software{Astropy \citep{astropy:2013, astropy:2018, astropy:2022},
Astroquery \citep{Ginsburg2019},
CASA \citep{McMullin2007},
lmfit \citep{Newville2016},
vis-sample \citep{Loomis2018}
}


\appendix
\input{obsAppendix}
\input{mapsAppendix}

\input{radAppendix}

\bibliography{3I}{}
\bibliographystyle{aasjournalv7}



\end{document}

%% file: Abstract.tex
\begin{abstract}

We report the detection of methanol (CH$_3$OH) toward interstellar comet 3I/ATLAS using the Atacama Compact Array of the Atacama Large Millimeter/Submillimeter Array (ALMA) on UT 2025 August 28, September 18 and 22, and October 1, and of hydrogen cyanide (HCN) on September 12 and 15. These observations spanned pre-perihelion heliocentric distances (\rh{}) of 2.6 -- 1.7 au. The molecules showed outgassing patterns distinct from one another, with HCN production being depleted in the sunward hemisphere of the coma, whereas CH$_3$OH was enhanced in that direction. Statistical analysis of molecular scale lengths in 3I/ATLAS indicated that CH$_3$OH included production from coma sources at $L_p>258$ km at 99\% confidence. However low signal-to-noise on long baselines, which sample emission on small spatial scales closest to the nucleus, prevented definitively ruling out CH$_3$OH as purely a parent species. In contrast, HCN production was indistinguishable from direct nucleus sublimation. The CH$_3$OH production rate increased sharply from August through October, including an uptick near the inner edge of the H$_2$O sublimation zone at \rh{} = 2 au. Compared to comets measured to date at radio wavelengths, the derived CH$_3$OH/HCN ratios in 3I/ATLAS of $124^{+30}_{-34}$ and $79^{+11}_{-14}$ on September 12 and 15, respectively, are among the most enriched values measured in any comet, surpassed only by anomalous solar system comet C/2016 R2 (PanSTARRS). 

\end{abstract}

%% file: body.tex
\section{Introduction} \label{sec:intro}
%Comets provide a window into the early solar system. They formed in the cold disk midplane of the protosolar disk during the era of planet formation and were subsequently scattered to the Kuiper Belt, the Scattered Kuiper Disk, or the Oort cloud. Cryogenically preserved for the last $\sim$4.5 Gyr in the  cold outer solar system, comets may serve as ``fossils'' of solar system formation, with the volatile composition of their nuclei reflecting the chemistry and prevailing conditions present where and when they formed \citep{Bockelee2004,Mumma2011a,Bockelee2017}.  

Interstellar objects afford an opportunity to understand planet formation around another star, transporting a record of their protoplanetary disk for study as they pass through the solar system. To date, only three clearly interstellar objects have been found. The first, 1I/'Oumuamua, appeared asteroidal, with no discernible coma despite the detection of non-gravitational accelerations \citep{Michelli2018}. The first interstellar comet, 2I/Borisov, displayed an unusual CO/HCN ratio \citep{Cordiner2020} despite a CN/H$_2$O ratio consistent with solar system comets \citep{McKay2020a}. 

The recent discovery of 3I/ATLAS \citep{Seligman2025} and subsequent studies have revealed the composition of an interstellar object in unprecedented detail. A recurring theme in these studies has been the remarkable composition of 3I/ATLAS compared to solar system comets. For instance, JWST observations at heliocentric distance \rh{} = 3.32 au found a CO$_2$-dominated coma with CO$_2$/H$_2$O = $7.6\pm0.3$ and CO$_2$/CO = $4.6\pm0.2$ \citep{Cordiner2025a}. Such a high CO$_2$/H$_2$O ratio was interpreted as a sign that the nucleus was exposed to higher levels of radiation than solar system comets, or instead may have formed close to the CO$_2$ snowline in the disk. Alternatively, it may be a signature of sublimation from ices heavily affected by galactic cosmic ray processing \citep{Maggiolo2025}. Similarly, CN, Ni, and Fe were measured in the optical, with an extremely enriched NiI/FeI abundance ratio found at \rh{} = 2.64 au \citep{Hoogendam2025,Hutsemekers2025}. At longer wavelengths, JCMT detected HCN at \rh{} = 2.1 au with an enriched abundance relative to H$_2$O of 0.2\%. Alongside these remarkable findings, the increase in H$_2$O outgassing was also followed through its photodissociation product, OH. The OH ($\mathrm{A}^2\Sigma-\mathrm{X}^2\Pi$) band near 3085 \AA \ was detected at ultraviolet wavelengths near \rh{} = 2.9 au (along with upper limits at \rh{} = 4.37 au) and OH 18 cm emission was seen at radio wavelengths at \rh{} = 1.43 au \citep{Xing2025,Alvarez2025,Crovisier2025}. 

Here we detail serial observations of 3I/ATLAS with the ALMA Atacama Compact Array and the detection of CH$_3$OH and HCN. These observations sampled multiple CH$_3$OH transitions centered near 338 GHz and 350 GHz pre-perihelion from \rh{} = 2.6 - 1.7 au, and the HCN (\Ju{} = $4-3$) transition at \rh{} = 2.17 and 2.08 au. We report molecular production rates and kinematics, parent scale lengths, coma radial temperature profiles, and integrated flux maps of molecular emission. Section~\ref{sec:obs} provides a description of the observations. Section~\ref{sec:results} presents our results from these data. Section~\ref{sec:discussion} discusses the kinematics and origins of CH$_3$OH and HCN in 3I/ATLAS, and places this interstellar object into context with solar system comets measured to date.

\section{Observations and Data Reduction} \label{sec:obs}
Comet 3I/ATLAS (hereafter 3I) reached perihelion ($q=1.35$ au) on 2025 October 29. We conducted pre-perihelion observations toward 3I on UT 2025 August 28, September 12, 15, 18, and 22, and October 1 (DDT 2024.A.00049.S, PI Roth; 2024.1.00477.T and 2024.1.00137.T, PI Cordiner) during Cycle 11 using the ALMA Atacama Compact Array (ACA) with the Band 7 receiver, covering frequencies between 335.5 and 354.6 GHz ($\lambda$ = 0.84 -- 0.89 mm) in six non-contiguous spectral windows. The details of our observing, calibration, and imaging procedure are provided in Appendix~\ref{sec:obslog} along with an observing log. 

\section{Results and Analysis} \label{sec:results}
\begin{figure*}
\gridline{\fig{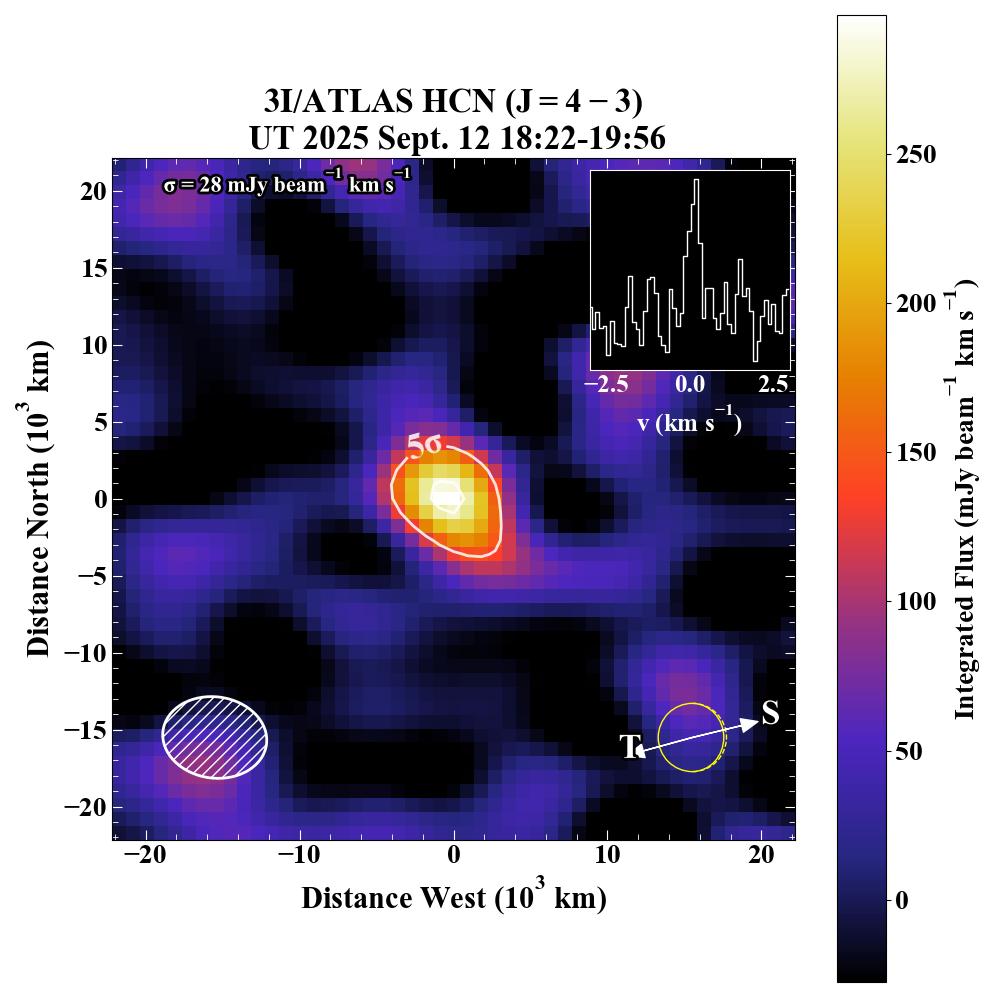}{0.45\textwidth}{(A)}
          \fig{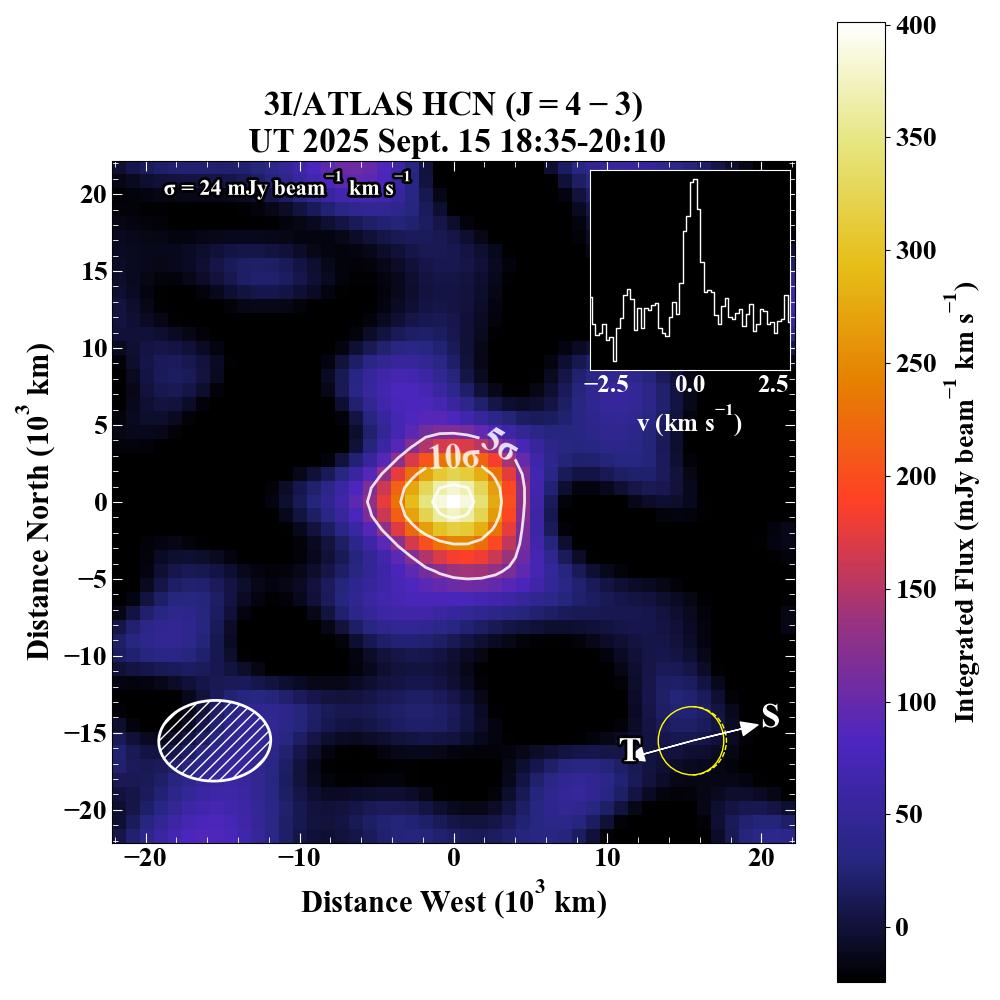}{0.45\textwidth}{(B)}
}
\gridline{\fig{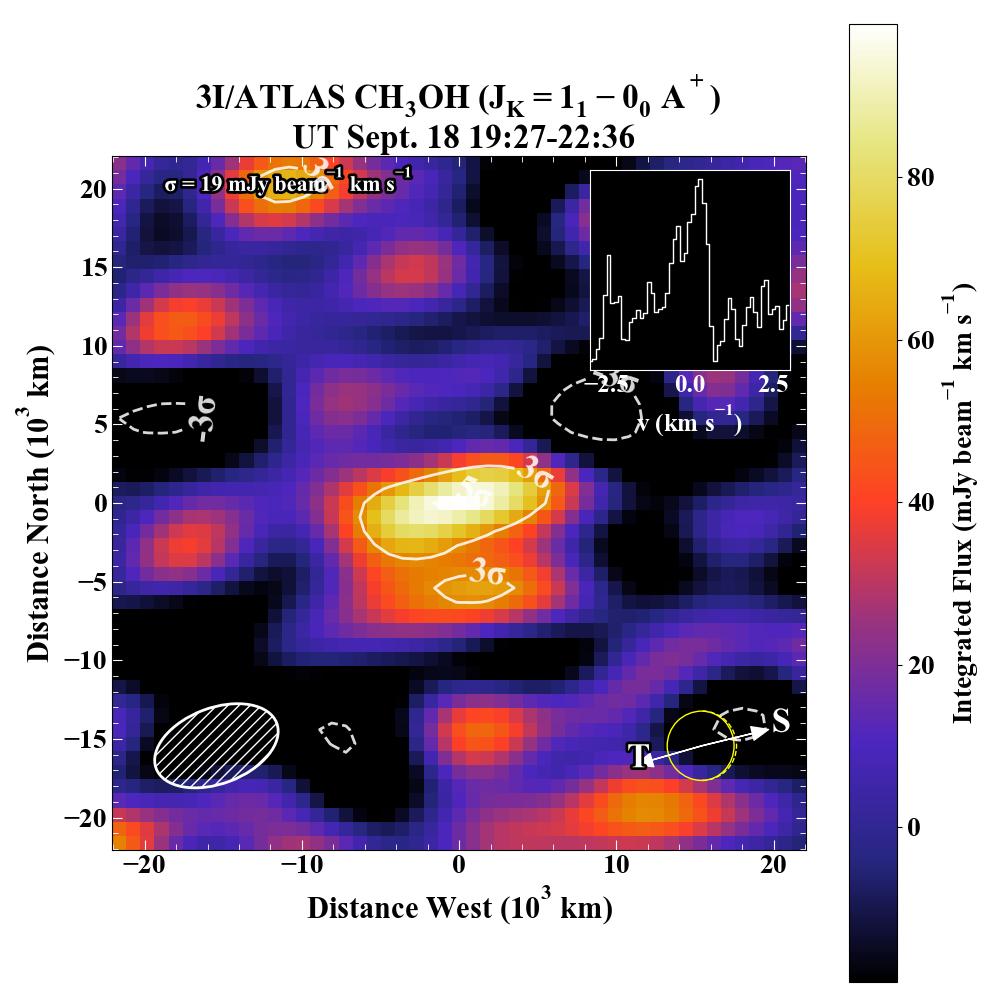}{0.45\textwidth}{(C)}
          \fig{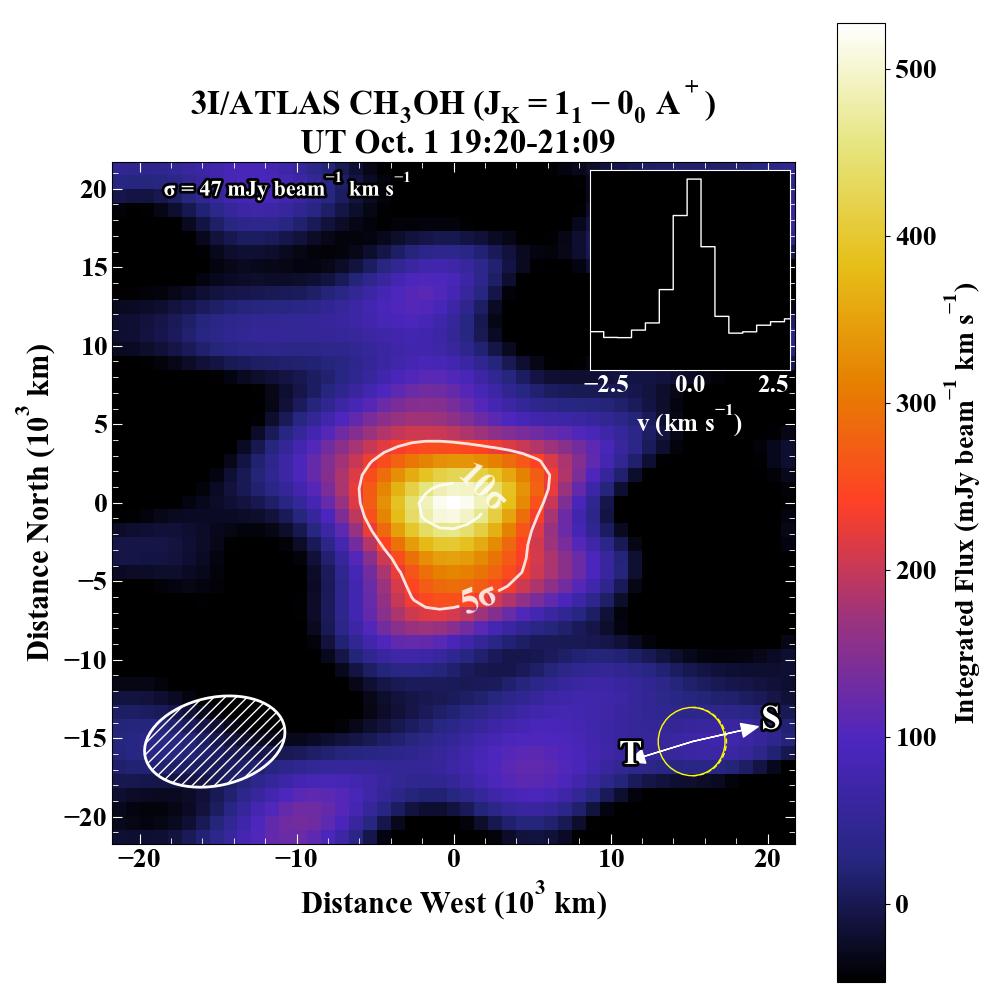}{0.45\textwidth}{(D)}
}
\caption{\textbf{(A)--(D).} Spectrally integrated flux maps for HCN on September 12 and 15 and for CH$_3$OH on September 18 and October 1. Contour intervals in each map are given in multiples of the rms noise.  The rms noise ($\sigma$, mJy beam$^{-1}$ km s$^{-1}$) is indicated in the upper left corner of each panel. Contours are $5\sigma$ and $10\sigma$ for HCN and for CH$_3$OH on October 1, and $3\sigma$ and $5\sigma$ for CH$_3$OH on September 18. Sizes and orientations of the synthesized beam (Table~\ref{tab:obslog}) are indicated in the lower left corner of each panel. The comet's observer-centered illumination ($\phi_{\mathrm{STO}} \sim$ 20$\degr$), as well as the direction of the Sun and dust trail, are indicated in the lower right. A spectrum of the HCN (\Ju{} = $4-3$) transition from a $10\arcsec$ diameter aperture centered on the peak emission is shown in the upper right. The CH$_3$OH map on September 18 shows the same extract for the high resolution spectrum of the $J_K=1_1-0_0 A^+$ transition near 350 GHz, whereas the October 1 map shows this transition at low spectral resolution on October 1.
\label{fig:main-maps}}
\end{figure*}

We detected molecular emission from CH$_3$OH on August 28, September 18 and 22, and October 1, and from HCN on September 12 and 15. These detections included up to thirteen lines of CH$_3$OH spanning a wide range of excitation energies ($E_u$ = 16.8 K -- 114.8 K), consisting of the $J_K=7_K-6_K$ ladder near 338 GHz and the $J_K=1_1-0_0 A^+$ and $J_K=4_0-3_{-1}E$ transitions near 350.905 and 350.687 GHz, respectively. HCN and CH$_3$OH emission were modestly spatially resolved at the $5\sigma$ level. Spectrally integrated flux maps for each species are shown in Figure~\ref{fig:main-maps}. Additional maps, as well as a list of all detected molecular transitions and their line parameters, are provided in Appendix \ref{sec:extraMaps}.

\subsection{Radiative Transfer Modeling}\label{subsec:radiative}
We modeled molecular line emission using the SUBLIME radiative transfer code for cometary atmospheres \citep{Cordiner2022}. SUBLIME includes a full non-LTE treatment of coma gases, including collisions with H$_2$O and electrons and pumping by solar radiation, along with a time-dependent integration of the energy level population equations. We used explicitly calculated HCN-H$_2$O collisional rates \citep{Zoltowski2025}. There are no published state-to-state CH$_3$OH-H$_2$O collisional rates, so we calculated them using the thermalization approximation \citep{Crovisier1987,Biver1999,Bockelee-Morvan2012}, using an average collisional cross-section with H$_2$O of $5\times10^{-14}$ cm$^{-2}$. Solar photodissociation rates were adopted from \cite{Hrodmarsson2023}. In the absence of contemporaneously measured $Q$(H$_2$O) values, we scaled the $Q$(OH)=$(5.7\pm0.6)\times10^{28}$ \ps{} measured by \cite{Crovisier2025} at \rh{} = 1.43 au to the \rh{} of the ACA observations assuming insolation-driven (\rh{}$^{-2}$) dependence and an \ce{H2O -> OH + H} branching ratio of 85.5\% \citep{Huebner2015}. We used these for setting the activity level and density of our coma radiative transfer models. The assumed values were calculated as: 

\begin{equation}\label{eq:oh}
Q(\mathrm{H_2O})(r_\mathrm{H}) = \frac{Q(\mathrm{OH})(r_\mathrm{H}=1.43 \mathrm{au})}{85.5\%}  \left(\frac{1.43 \mathrm{au}}{r_\mathrm{H}}\right)^2
\end{equation}

\begin{deluxetable*}{cccccccc}
\tablenum{1}
\tablecaption{Molecular Kinematics and Abundances in 3I/ATLAS\label{tab:kinematics}}
\tablewidth{0pt}
\tablehead{
\colhead{Date} & \colhead{UT Time} & \colhead{$v_1^{(a)}$} & \colhead{$v_2^{(b)}$} & \colhead{$Q_1/Q_2^{(c)}$} & \colhead{$T_\mathrm{kin}^{(d)}$} & \colhead{$L_p^{(e)}$} & \colhead{$Q_x^{(f)}$} \\
\colhead{(2025)} & \colhead{} & \colhead{(\kms{})} & \colhead{(\kms{})} & \colhead{ } & \colhead{(K)} & \colhead{(km)} & \colhead{($10^{26}$ \ps{})} 
}
\startdata
\hline
\multicolumn{8}{c}{HCN and CH$_3$OH 3D Models} \\
\hline
\multicolumn{8}{c}{HCN} \\
Sept. 12 & 18:22-19:56 & $<0.02$ & $0.35\pm0.05$ & $0.04\pm0.01$ & (45) & (0) & $0.04\pm0.01$ \\
Sept. 15 & 18:35-20:10 & $0.31\pm0.05$ & $0.24\pm0.02$ & $0.7\pm0.1$ & (45) & $<1430$ & $0.09\pm0.01$ \\
\hline
\multicolumn{8}{c}{CH$_3$OH} \\
Sept. 18 & 19:27-22:36 & $0.51\pm0.04$ & $0.31\pm0.03$ &  $1.6\pm0.5$ & (49) & $4107^{+2621}_{-1820}$ & $8\pm1$ \\
\hline
\hline
\multicolumn{8}{c}{HCN and CH$_3$OH 1D Models} \\
\hline
 Date & UT Time & $v^{(g)}$ & $\Delta v^{(h)}$ & & $T_\mathrm{kin}$ & $L_p$ & $Q_x$   \\
 (2025) & & (\kms{}) & (\kms{}) & & (K) & (km) & ($10^{26}$ \ps{})   \\
 \hline
 \multicolumn{8}{c}{HCN} \\
%Sept. 12, 15 & 18:22-20:10 & $0.27\pm0.03$ & $0.24\pm0.02$ & $0.5\pm0.1$ &  (40) & $<$1114 (3$\sigma$) & $0.10\pm0.01$ \\
Sept. 12 & 18:22-19:56 & $0.19\pm0.03$ & $0.11\pm0.02$ & & (45) & $<410$ & $0.05\pm0.01$  \\
Sept. 15 & 18:35-20:10 & $0.23\pm0.02$ & $0.05\pm0.01$ & & (45) & $922^{+1023}_{-783}$ & $0.10\pm0.01$  \\
\hline
 \multicolumn{8}{c}{CH$_3$OH} \\
Aug. 28  & 20:59-22:35 & (0.37) & (0) & & (40) & (0) & $5.0\pm0.6$   \\
Sept. 18 & 17:07-18:34 & (0.44) & (0) &  & $49\pm12$ & (2884) & $9.6\pm0.7$   \\
Sept. 18 & 19:27-22:36 & $0.44\pm0.03$ & $-0.13\pm0.04$ & & (49) & $2884^{+1860}_{-1192}$ & $9\pm1$  \\ 
Sept. 22 & 16:34-18:13 & (0.43) & (0) & & $52\pm4^{**}$ & (0) & $12.1\pm0.5$ \\
Oct. 1   & 19:20-21:09 & (0.46) & (0) & & $50\pm2^{**}$ & (0) & $22.9\pm0.4$  \\
\enddata 
\tablecomments{\sups{a} Gas expansion speed in the sunward hemisphere $(R_1)$. \sups{b} Gas expansion speed in the anti-sunward hemisphere $(R_2)$. \sups{c} Ratio of production rates in the sunward vs.\ anti-sunward hemispheres $(R_1/R_2)$. \sups{d} Gas kinetic temperature. \sups{e} Parent scale length. \sups{f} Molecular production rate. \sups{g} Gas expansion speed. \sups{h} Global model Doppler offset.  $^{**}$Inner temperature point from the best-fit radial temperature profile (Appendix~\ref{sec:fourier}). All values in parentheses are assumed.
}
\end{deluxetable*}

Observations of CH$_3$OH on August 28, September 18 and 22, and October 1 used a relatively low spectral resolution (0.87 \kms{}; Figure~\ref{fig:main-maps}D). The intrinsic width of the emission lines is directly related to the gas expansion speed; thus, when the lines are sufficiently spectrally resolved, detailed kinematics can be retrieved from the line shape. For the low spectral resolution observations, the line full width at half maximum (FWHM) was comparable to the instrumental resolution; therefore, we could not retrieve robust kinematics directly from fits to the lines. However, additional observations of CH$_3$OH on September 18 using higher spectral resolution (0.21 \kms{}) revealed a well-resolved and asymmetric line profile, inconsistent with uniform outflow (Figure~\ref{fig:main-maps}C). We used the full three-dimensional version of SUBLIME to model these data following methods in previous studies \citep{Cordiner2022,Roth2023} and dividing the coma into two outgassing regions, $R_1$ and $R_2$, each with independent molecular production rates ($Q_1$,$Q_2$), constant gas expansion velocities ($v_1$,$v_2$), and parent scale lengths ($L_{p1},L_{p2}$). $L_p$ parameterizes the distance from the nucleus at which a molecule formed ($L_p$ = 0 km is direct nucleus release) and is related to the parent photodissociation rate ($\beta_p$, \ps{}) as $L_p = v/\beta_p$. This is the simplest model capable of fitting the observed asymmetric CH$_3$OH line profile. 

For consistency, we also applied a 1D version of SUBLIME \citep{Cordiner2024} to the high resolution CH$_3$OH spectra and compared against our results from the 3D model. In addition to $Q,v,$ and $L_p$, the 1D model includes a global Doppler offset, $\Delta v$, as a free parameter. For a comet with well-constrained astrometry such as 3I ($3\sigma$ uncertainties on $d\Delta/dt\sim0.1$ \ms{}), a non-zero $\Delta v$ effectively creates a 1D model where $v$ is an average over the expansion speeds in the sunward and anti-sunward hemispheres rather than a spherically symmetric model. In the absence of firm constraints on kinematics for the remaining CH$_3$OH observations, we used a 1D model to analyze the low resolution data, extrapolating the expansion speed from the best-fit 1D models of the high resolution data and scaling with \rh{}. On September 18 and 22 and October 1, we also determined the coma kinetic temperature.

\begin{figure*}
\gridline{\fig{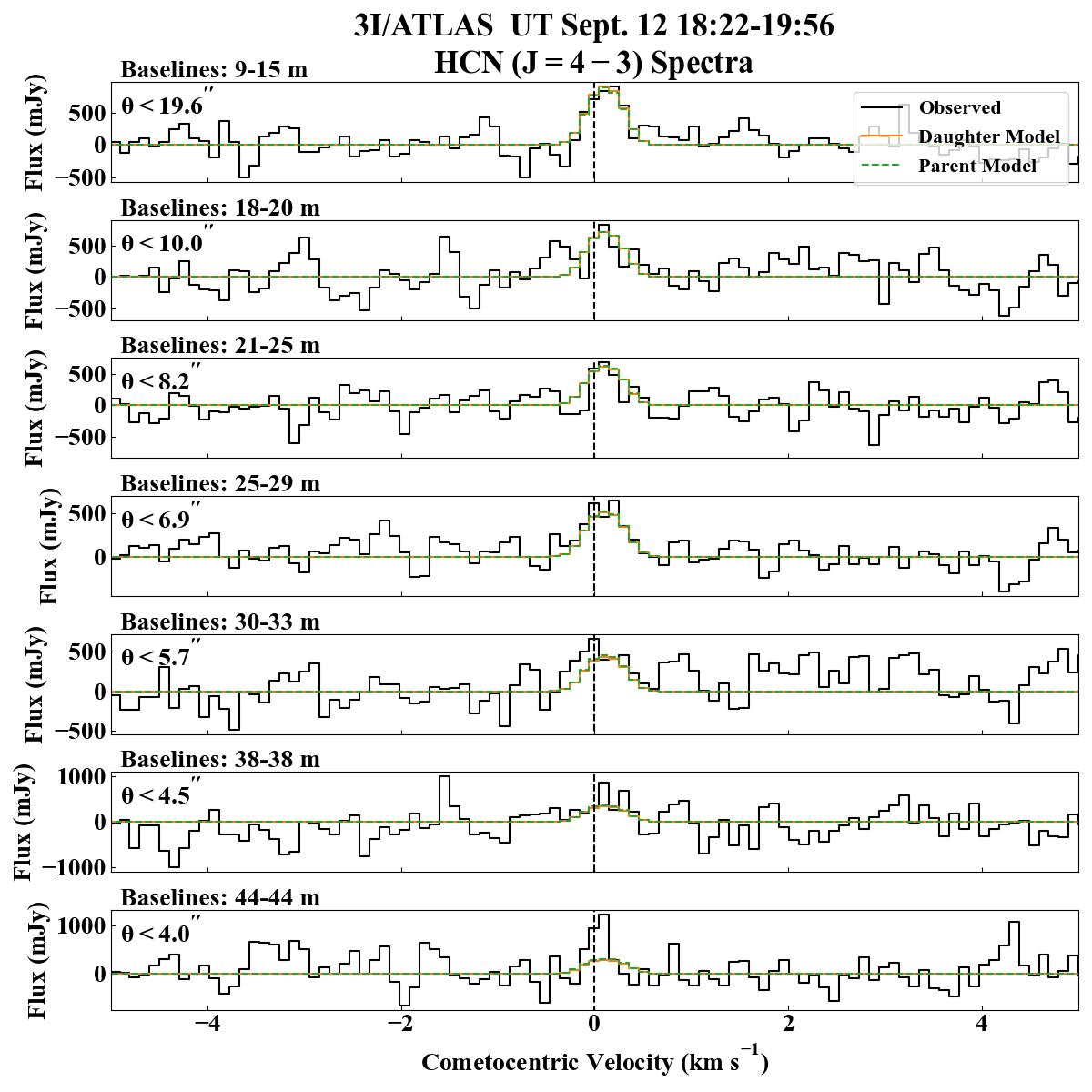}{0.45\textwidth}{(A)}
          \fig{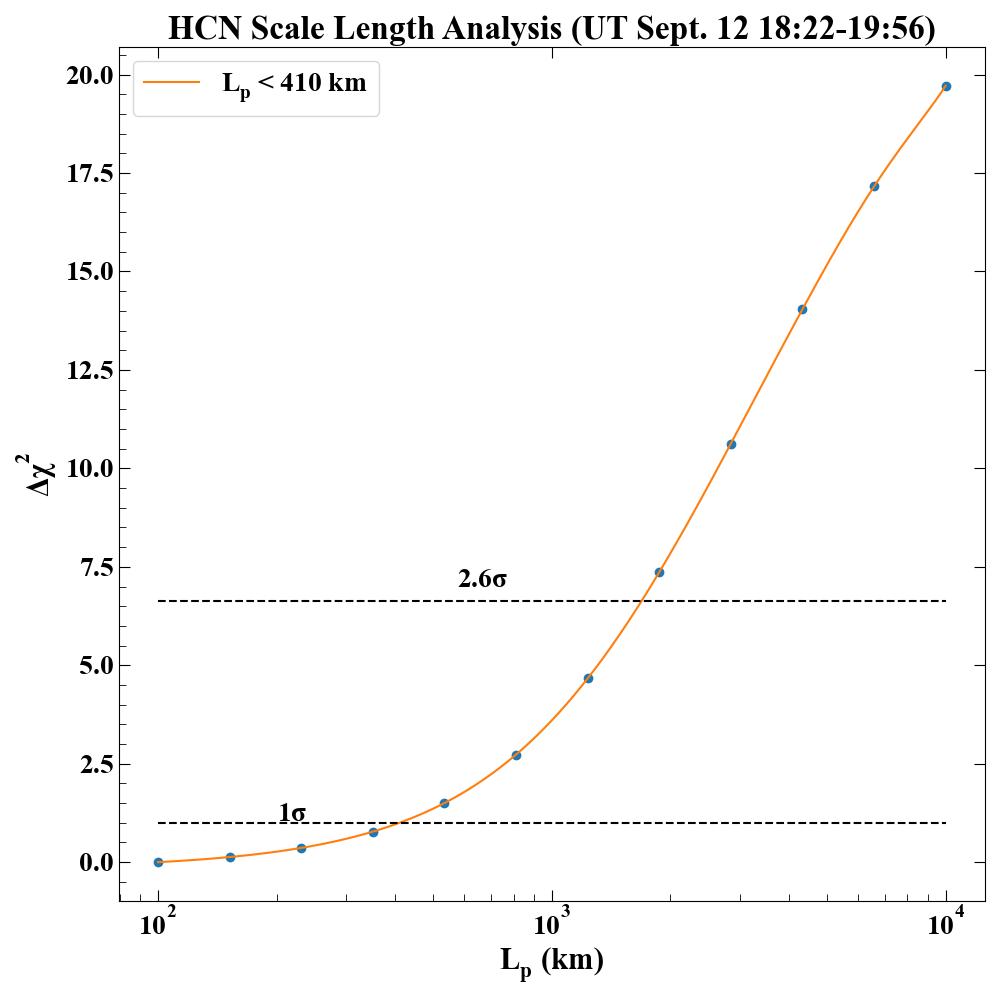}{0.45\textwidth}{(B)}
}
\gridline{\fig{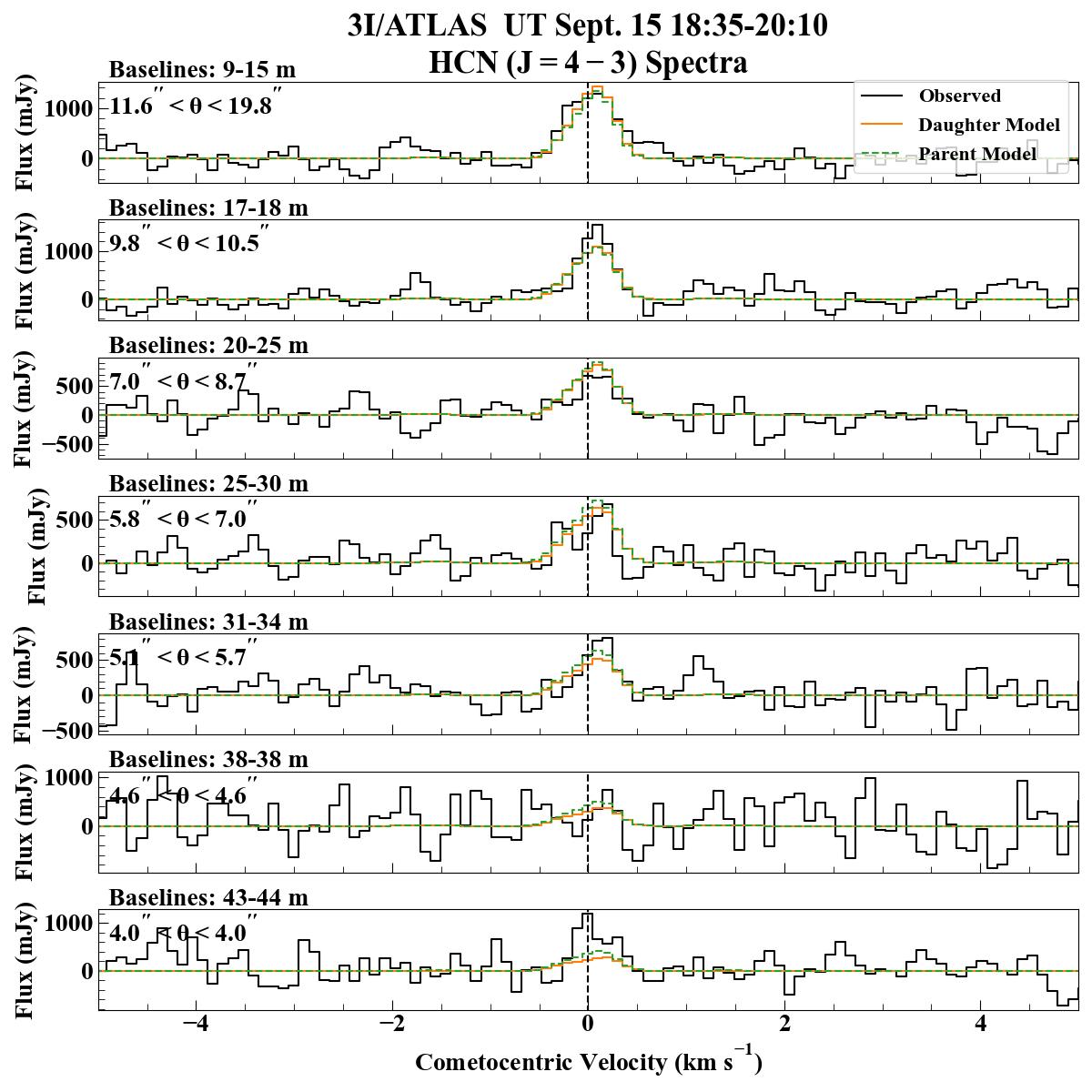}{0.45\textwidth}{(C)}
          \fig{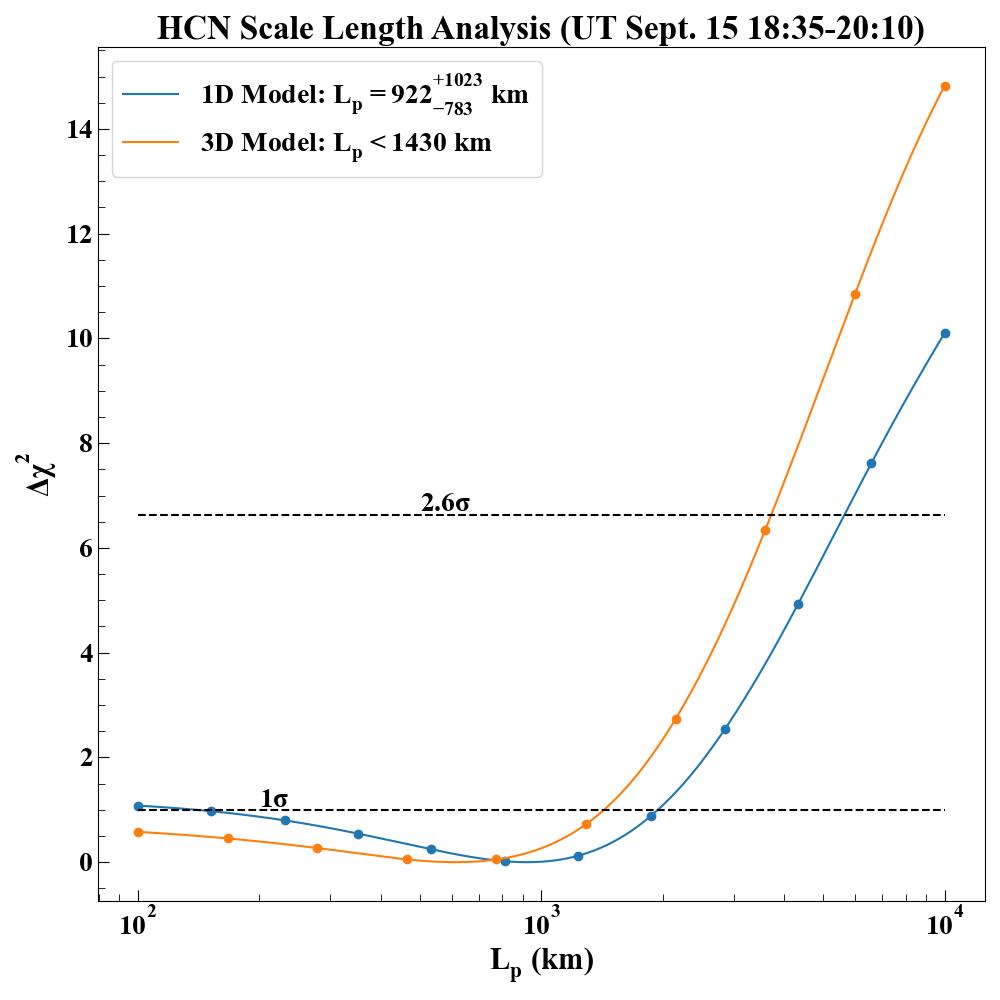}{0.45\textwidth}{(D)}
}
\caption{\textbf{(A).} HCN (\Ju{} = $4-3$) spectrum in 3I on Sept. 12, with each panel representing spectra extracted from differing baseline ranges (angular scales). The spectra are displayed with a frequency resolution of 122 kHz (velocity resolution 0.13 \kms{}). The best-fit 1D daughter and parent models are overplotted. A dashed vertical line at $v=0$ \kms{} emphasizes the redward offset of the line center. \textbf{(B).} $\Delta\chi^2$ analysis for the HCN $L_p$ on September 12 (Appendix~\ref{sec:fourier}). \textbf{(C--D).} Plots for HCN (\Ju{} = $4-3$) in 3I on September 15, with traces and labels as in panels (A) and (B).
\label{fig:hires-models}}
\end{figure*}

HCN was also observed at high (0.13 \kms{}) resolution, sufficient to resolve the line. Similar to CH$_3$OH, we compared fits from both the 3D and 1D versions of SUBLIME. The HCN line profiles on September 12 and 15 both showed a clear redward offset of the line center, although there are differences in the line profiles as a function of baseline between the two dates. In all cases, we performed parameter optimization via least-squares fitting in the Fourier domain, minimizing the residual for the observed and modeled visibility amplitudes. A full description of our Fourier-domain modeling formalism is provided in Appendix~\ref{sec:fourier}. Our results and best-fit models are summarized in Table~\ref{tab:kinematics}. Figure~\ref{fig:hires-models} shows a model-data comparison for HCN, and Figure~\ref{fig:ch3oh-models} shows the same for the high-resolution CH$_3$OH spectra on September 18. Appendix~\ref{sec:fourier} contains the same for the low resolution data, along with retrieved radial temperature profiles for September 22 and October 1.

\section{Discussion and Interpretation} \label{sec:discussion}

\subsection{Complex Line Profile and Production Mechanisms of HCN}\label{subsec:hcn-kinematics}
Our measurements of HCN and CH$_3$OH in 3I reveal quite different outgassing patterns between the two species. HCN showed a significant redward shift of the line center on September 12 and 15, indicative of excess production of HCN in the projected anti-sunward direction of the coma. A red excess at $v\sim0.75$ \kms{} is seen on the short baselines on both dates (Figure~\ref{fig:hires-models}); however, it is present in the maps at only $3\sigma$ and $4\sigma$ confidence on September 12 and 15, respectively. We integrated the blue and red components of the HCN line on both dates to provide a comparison of the overall line asymmetry and how it evolved with time. Figure~\ref{fig:trend} shows blueward emission was detected at only $3\sigma$ confidence on September 12 and the red emission showed an extension to the south. The balance shifted on September 15, with considerably more blue emission but a remaining southern red extension. 

The 3D model has difficulty fitting the highly asymmetric line profile on September 12. The model is consistent with the red/blue split of the maps, with HCN production that was dominated by sources in the projected anti-sunward hemisphere (25 times higher production compared to the sunward side) and an expansion speed $v_2=0.35\pm0.05$ \kms{}. The nominal upper limit on $v_1<0.02$ \kms{} is consistent with this interpretation. However, the 3D model indicates that the anti-sunward hemisphere was only 1.4 times more productive than the sunward side on September 15, with $v_1=0.31\pm0.05$ \kms{} and $v_2=0.24\pm0.02$ \kms{}. \cite{Santana-Ros2025} found a nucleus rotational period of 16.16 hours for 3I, so our observations on September 12 and 15 (separated by 72 hours) would have sampled very different rotational phases. Thus, the evolution of the line profile and maps suggests that projection effects may have been important, potentially owing to changing orientations of the HCN active sites on the nucleus along the line of sight. The very low $v_1$ on September 12 therefore may not represent the true bulk HCN expansion speed.

The results from the 1D models are consistent with this interpretation. The best-fit $\Delta v$ was significantly more redward on September 12 along with a lower average $v$ compared to September 15, reflecting the changing asymmetry of the line. We performed a $\Delta\chi^2$ analysis of our radiative transfer models to determine the $1\sigma$ (68\% confidence) and $2.6\sigma$ (99\% confidence) bounds on $L_p$ and test for coma production of HCN (see Appendix~\ref{sec:fourier}). Our $1\sigma$ bounds are $L_p < 410$ km and $L_p=922^{+1023}_{-783}$ km on September 12 and 15, respectively, from the 1D models. We were unable to retrieve a well-constrained $L_p$ from the 3D model on September 12, but for September 15 the 3D model gives $L_p<1430$ km. The $Q$(HCN) on a given date are in good agreement between the 1D and 3D models, as are the $L_p$ values. At 99\% confidence, the overall results are indistinguishable from HCN production from direct nucleus sublimation, consistent with solar system comets previously measured with ALMA \citep[e.g.,][]{Cordiner2014,Bogelund2017,Roth2021a}. 

The 1D and 3D models underestimate the HCN line strength on the longest baselines (smallest angular scales) on both dates, and underestimate the width of the line on the shortest baselines on September 15. On the one hand, this could be indicative of a strong temperature gradient in the coma, with our isothermal model underestimating the strength of the HCN (\Ju{} = $4-3$) line in the warmer, inner coma. However, examination of Figure~\ref{fig:hires-models}C in particular shows an evolution of the line shape with baseline length, with the FWHM of the line varying (see Appendix~\ref{sec:fourier} for more details) and in general increasing with angular scale. Such behavior could be indicative of coma gas acceleration \citep[e.g.,][]{Lammerzahl1987,Cordiner2025b}, with smaller velocities near the nucleus resulting in increased line strengths on long baselines compared to the predictions of our constant velocity model. Incorporating such effects could affect the derived $Q$(HCN) by a few percent \citep{Cordiner2025b}. Determining the $v(r)$ law for 3I's coma is beyond the scope of this work and the subject of a future publication.

\begin{figure*}
\gridline{\fig{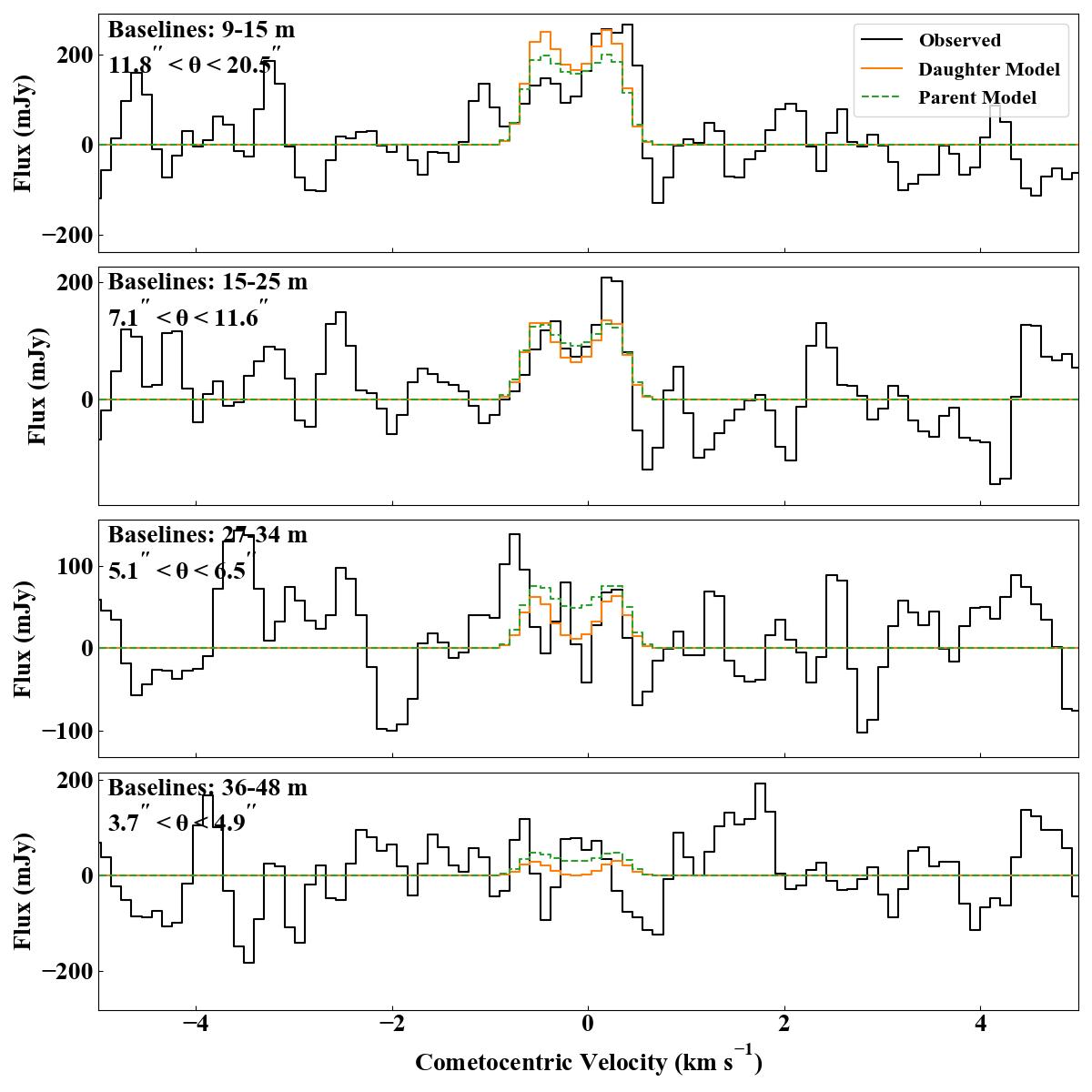}{0.45\textwidth}{(A)}
          \fig{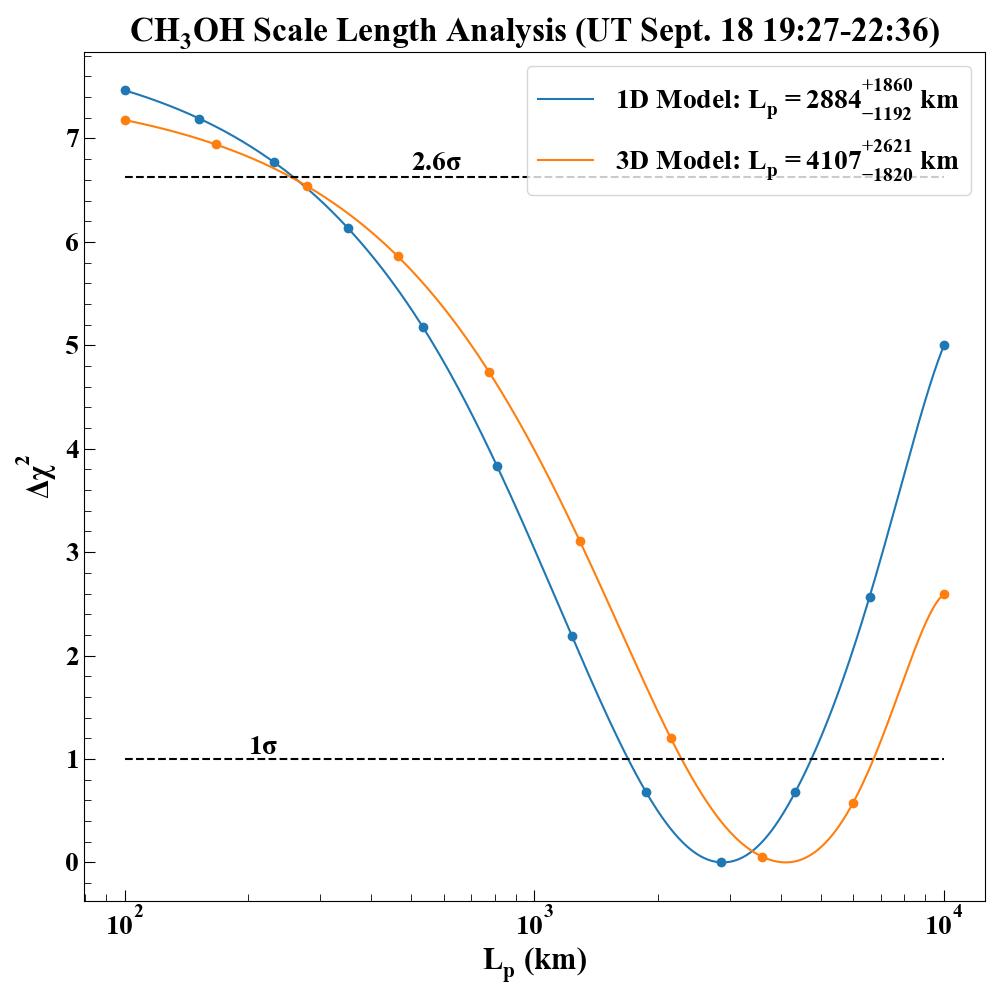}{0.45\textwidth}{(B)}
}
\gridline{\fig{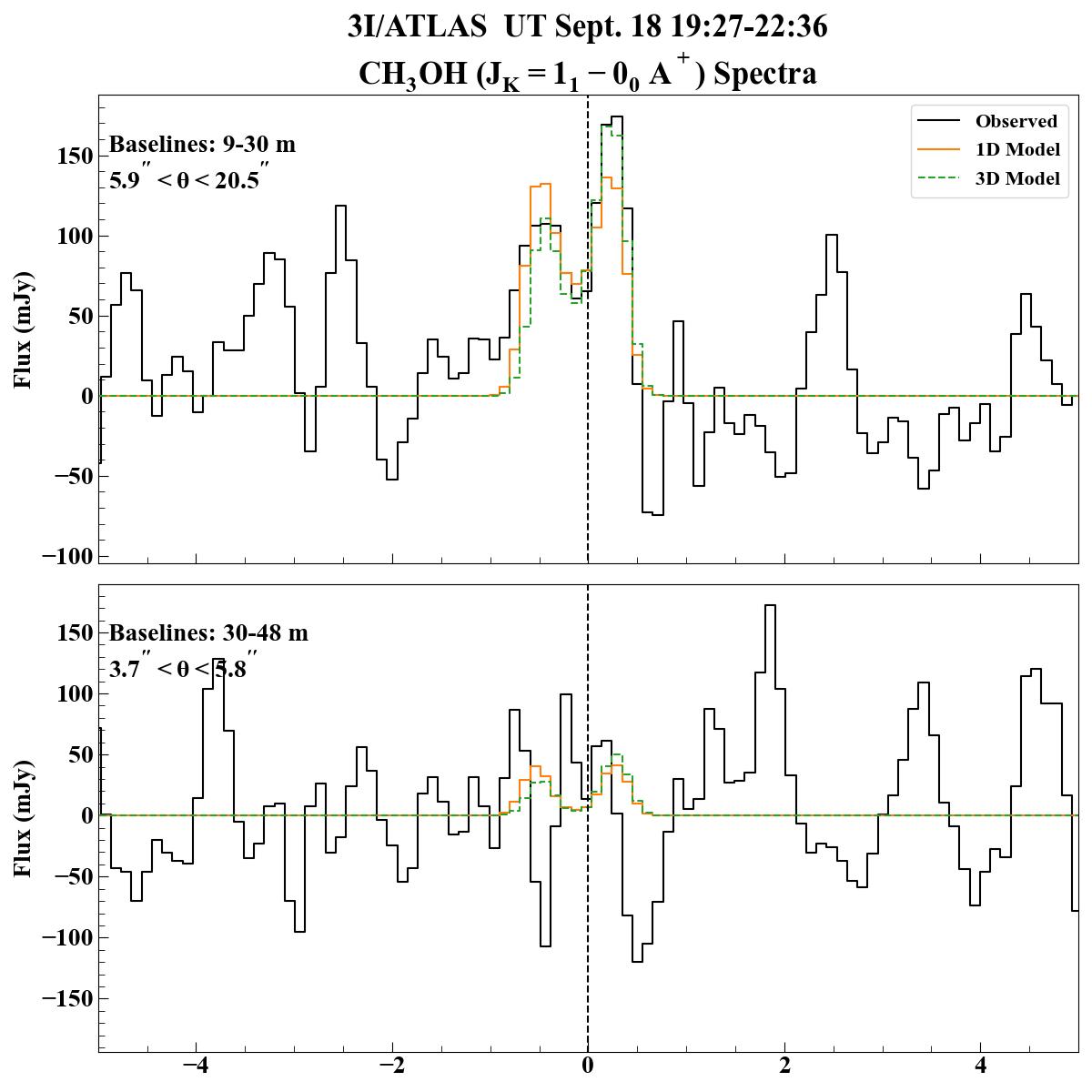}{0.45\textwidth}{(C)}
          \fig{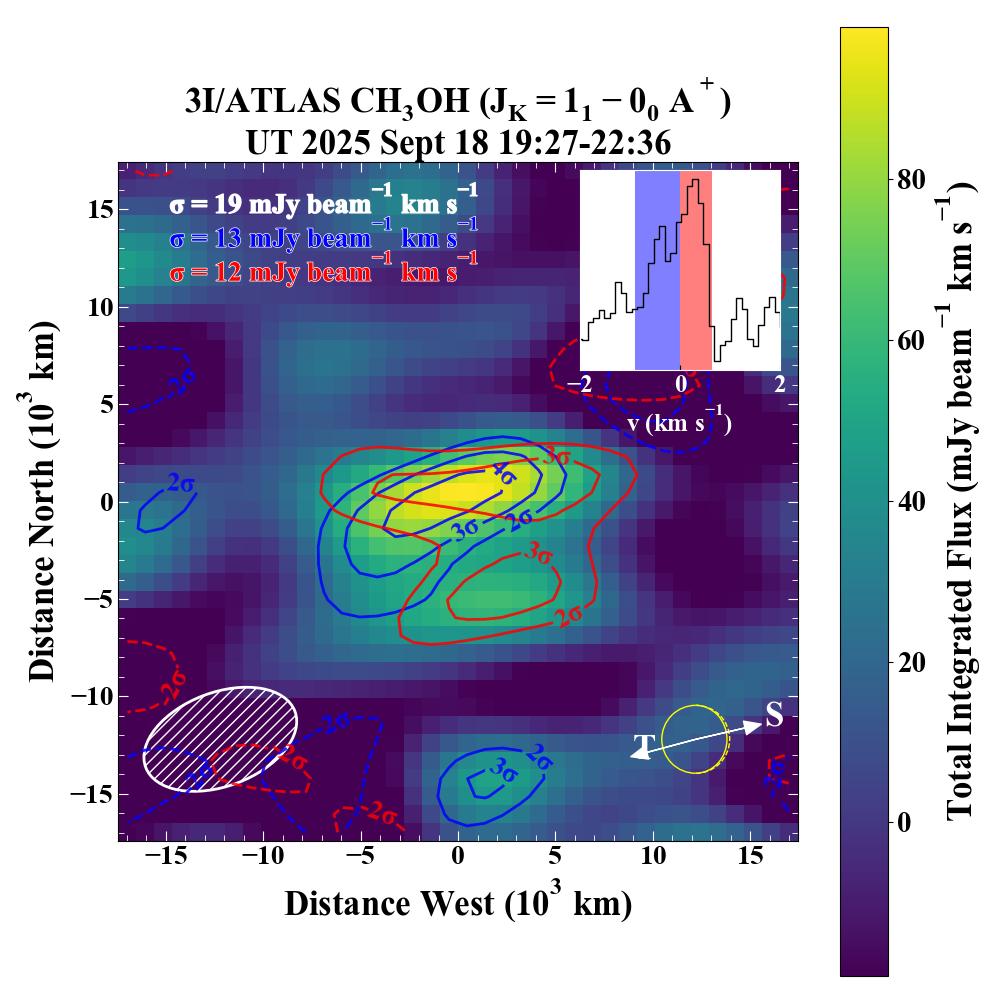}{0.45\textwidth}{(D)}
}
\caption{\textbf{(A).} CH$_3$OH ($J_K=1_1-0_0 A^+)$  spectrum in 3I on Sept. 18, with each panel representing spectra extracted from differing baseline ranges (angular scales). The spectra are displayed with a frequency resolution of 122 kHz (velocity resolution 0.21 \kms{}). The best-fit 1D daughter and parent models are overplotted. A dashed vertical line at $v=0$ \kms{} emphasizes the blueward offset of the line center. \textbf{(B).} $\Delta\chi^2$ analysis for the CH$_3$OH $L_p$ on September 18 from the 1D and 3D models (Appendix~\ref{sec:fourier}). \textbf{(C).} As in panel (A), but with a comparison of the best-fit 1D  and 3D  SUBLIME daughter models overplotted. Spectra are summed over a larger number of short baselines to emphasize the signal-to-noise ratio of the asymmetric line profile. \textbf{(D).} Integrated CH$_3$OH ($J_K=1_1-0_0A^+$) flux map on September 18, with contours for the red and blue components overlaid separately. Contours are in $1\sigma$ increments, with $2\sigma$ being the lowest contour level. The extracted spectral line profile shows the integration region for the blue and red maps across the line.
\label{fig:ch3oh-models}}
\end{figure*}

\subsection{Production Mechanisms and Kinematics of CH$_3$OH}\label{subsec:methanol-kinematics}
The CH$_3$OH results tell a different story. There was a blueward offset to the line center in contrast to HCN, and the line profile appears asymmetric in the maps (Figure~\ref{fig:main-maps}C) and when summed over the short--mid baseline lengths (Figure~\ref{fig:ch3oh-maps2}C). The 3D model returns a nominally higher $L_p=4107^{+2621}_{-1820}$ km than the $L_p=2884^{+1860}_{-1192}$ km of the 1D model, but the two agree within $1\sigma$. Despite appearing to better reproduce the asymmetry of the red and blue peaks on short baselines (Figure~\ref{fig:ch3oh-models}C), the 3D model does not provide a significantly better fit than the 1D model at the 95\% confidence level (Appendix~\ref{sec:fourier}). On the other hand, at 99\% confidence, both models agree that there was some coma production of CH$_3$OH at $L_p>258$ km. Interestingly, similar to HCN, the $J_K=1_1-0_0A^+$ maps on September 18 and October 1 (Figure~\ref{fig:main-maps}) show a southern extension of emission. Although the statistical significance is lower on September 18 (when there was sufficient spectral resolution to separate the blue and red components), the southern extension was associated with the red side of the line (as for HCN; Figure~\ref{fig:ch3oh-models}D).

However, the $J_K=7_K-6_K$ transitions do not show this southern extension (Appendix~\ref{sec:extraMaps}). Indeed, the $J_K=7_K-6_K$ transitions have relatively high excitation energies $(E_u>65$ K) and are likely populated closer to the nucleus, whereas the $J_K=1_1-0_0 A^+$ transition comes from lower energy levels and may probe larger distances in the coma. Thus, the latter should provide a robust test for extended source production, and it is possible that an active region on the nucleus in the projected southern hemisphere was important for extended sources. Our radial temperature profiles on September 22 and October 1 (Appendix~\ref{sec:fourier}) show that the coma stayed relatively warm out to large ($\sim13,000$ km) nucleocentric distance considering \rh{} ($\sim$2 au) during our observations. Taken together with the southern extension of the $J_K=1_1-0_0A^+$ line and the blue offset of the CH$_3$OH line center, these may be the signatures of CH$_3$OH production from icy grains in the projected sunward hemisphere alongside direct nucleus release elsewhere. However, Figure~\ref{fig:ch3oh-models} shows that CH$_3$OH was not firmly detected on the longest baselines, and we cannot completely rule out CH$_3$OH as a purely parent species. 

Overall, the significant differences in HCN and CH$_3$OH outgassing could be indicative of a nucleus with heterogeneous composition. Such behaviors have been observed in solar system comets. CH$_3$OH production from icy grains (also termed cometary halo ice primaries or CHIPs) has been reported previously in hyperactive comets such as 103P/Hartley 2 and 46P/Wirtanen \citep{Drahus2012,Coulson2020,Cordiner2023}, and the Rosetta mission to 67P/Churyumov-Gerasimenko demonstrated a chemically heterogeneous nucleus with two distinct hemispheres \citep[e.g.,][]{Luspay-Kuti2019,Herny2021} driving coma composition depending on the nucleus season. Alternatively, it is interesting to note that the CH$_3$OH and HCN line profiles were similar in the red (anti-sunward) hemisphere, yet HCN was depleted in the sunward hemisphere whereas CH$_3$OH was enriched. This is another line of evidence that the low HCN velocity in the sunward hemisphere on September 12 may have been a projection effect, with the illuminated side of the nucleus producing very little HCN and the derived $v_1$ not representative of the bulk HCN expansion speed toward the Sun. Additional observations of HCN are required to test this hypothesis.

\subsection{Comparison to Other Measurements of 3I/ATLAS}\label{subsec:compareK2}
Observations of 3I to date have detected a range of molecular species and tracers. Most relevant to this study are preliminary detections of HCN (\Ju{} = $3-2$) near 265 GHz with the JCMT \citep{Coulson2026} on September 7 and 14 and OH 18 cm emission with the Nan\c{c}ay radio telescope \citep{Crovisier2025} from October 13--19. 

When comparing the ACA and JCMT results, it is important to note that different lines were sampled and different model kinetic temperatures assumed (\Ju{} = $4-3$ with $E_u=42.5$ K and \tkin{} = 45 K for the ACA, \Ju{} = $3-2$ with $E_u=25.5$ K and \tkin{} = 35 K for JCMT). Nevertheless, in terms of kinematics and production rates, our HCN expansion speeds ($\sim$0.2 \kms{}) are lower than that reported in \cite{Coulson2026} ($v=0.46\pm0.14$ \kms{}) for September 14, although the two values agree at the $2\sigma$ level. Our $Q$(HCN) of $(5\pm1)\times10^{24}$ \ps{} and $(1.0\pm0.1)\times10^{25}$ \ps{} (derived from the 1D models) on September 12 and 15, respectively, can be compared to their $Q$(HCN) of $(1.3\pm0.4)\times10^{25}$ \ps{} and $(4.0\pm1.7)\times10^{25}$ \ps{} on September 7 and 14, respectively. On the one hand, these values could indicate significant variability in $Q$(HCN) on day-to-day time scales. 

Alternatively, the evidence for coma acceleration in our spectra (Figure~\ref{fig:hires-models}, Appendix~\ref{sec:fourier}) coupled with differences in beam size may reconcile the two datasets. The JCMT beam size of $18\farcs4$ at 265 GHz is comparable to the maximum recoverable scale of the ACA ($19\farcs6$) at 354 GHz, yet nearly three times the size of the synthesized beam. Their model (like ours) assumed constant $v$, and their derived value is consistent with what would be expected from the HCN line FWHM on the shortest ACA baselines. Indeed, fixing $v=0.46$ \kms{} and recalculating $Q$(HCN) on September 15 gives $(2.7\pm0.2)\times10^{25}$ \ps{}, in agreement with JCMT within $1\sigma$ uncertainty.

\subsection{Comparison to Solar System Comets}\label{subsec:compareSolar}
\begin{figure*}
\gridline{\fig{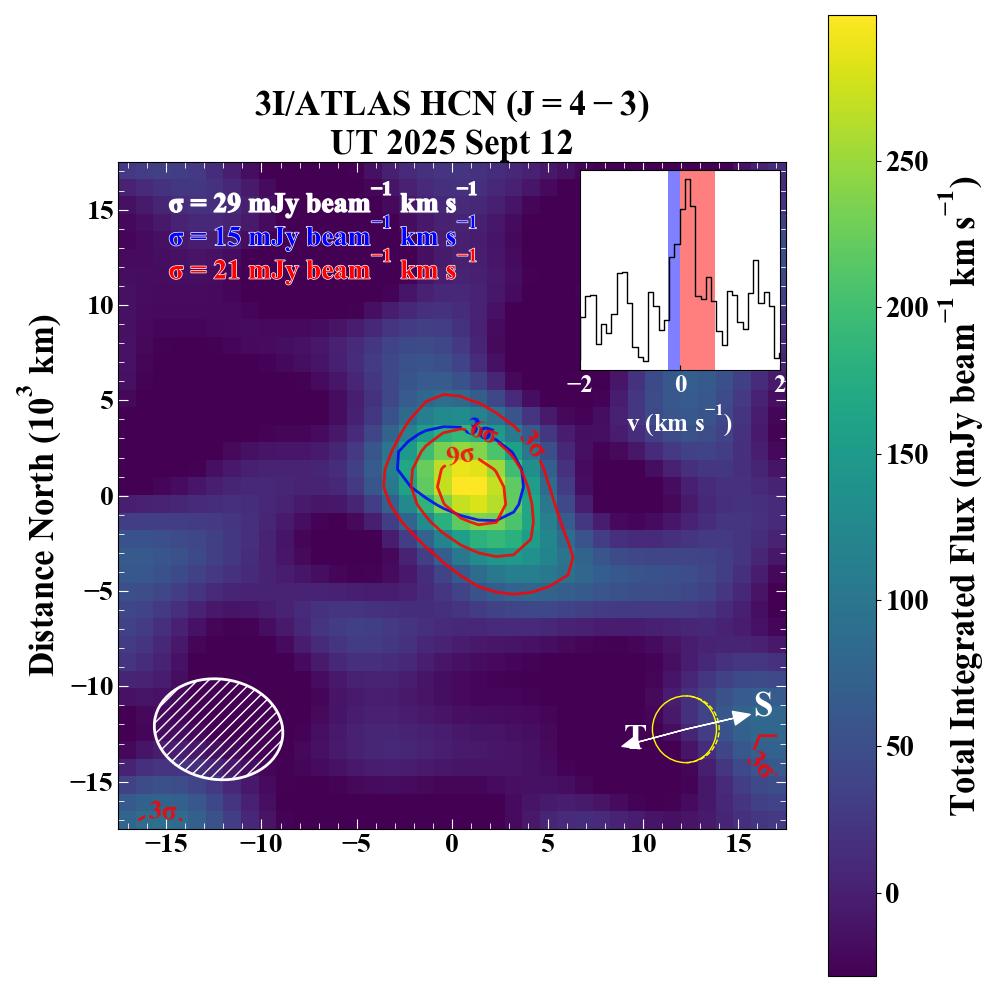}{0.45\textwidth}{(A)}
          \fig{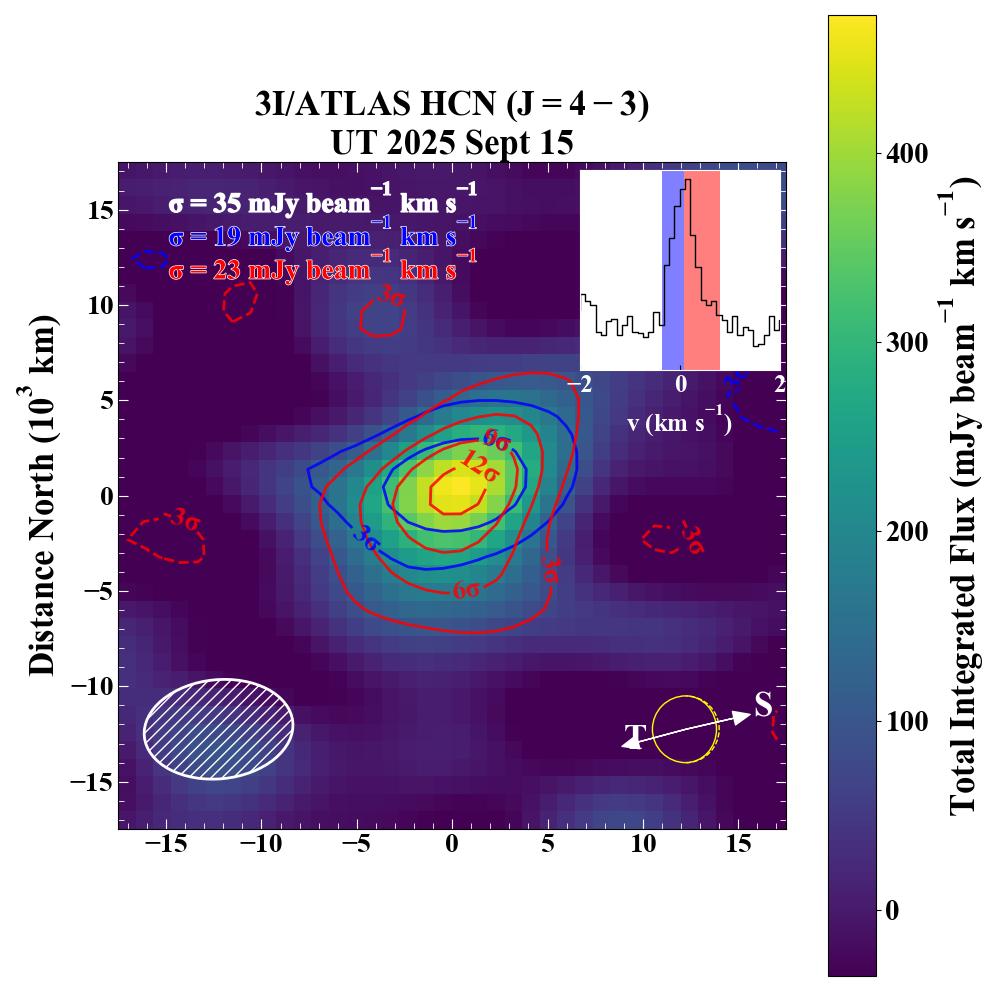}{0.45\textwidth}{(B)}
}
\gridline{\fig{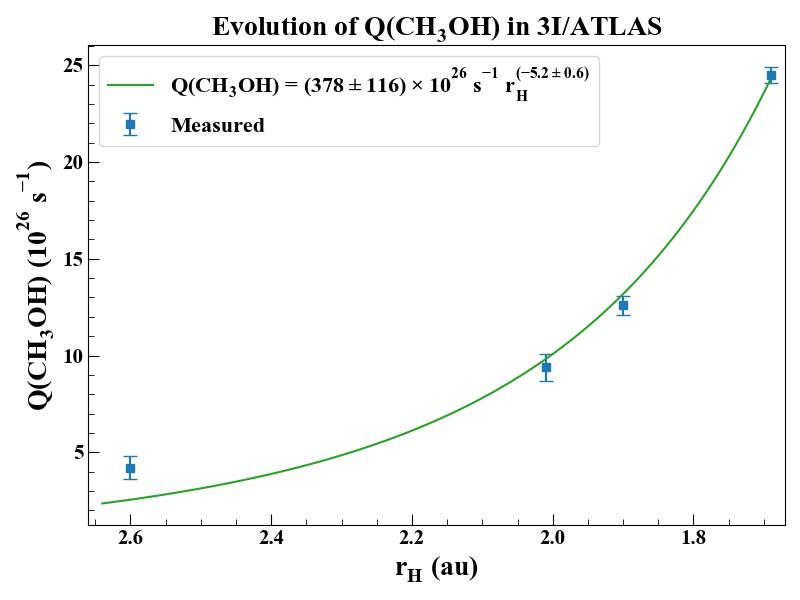}{0.45\textwidth}{(C)}
          \fig{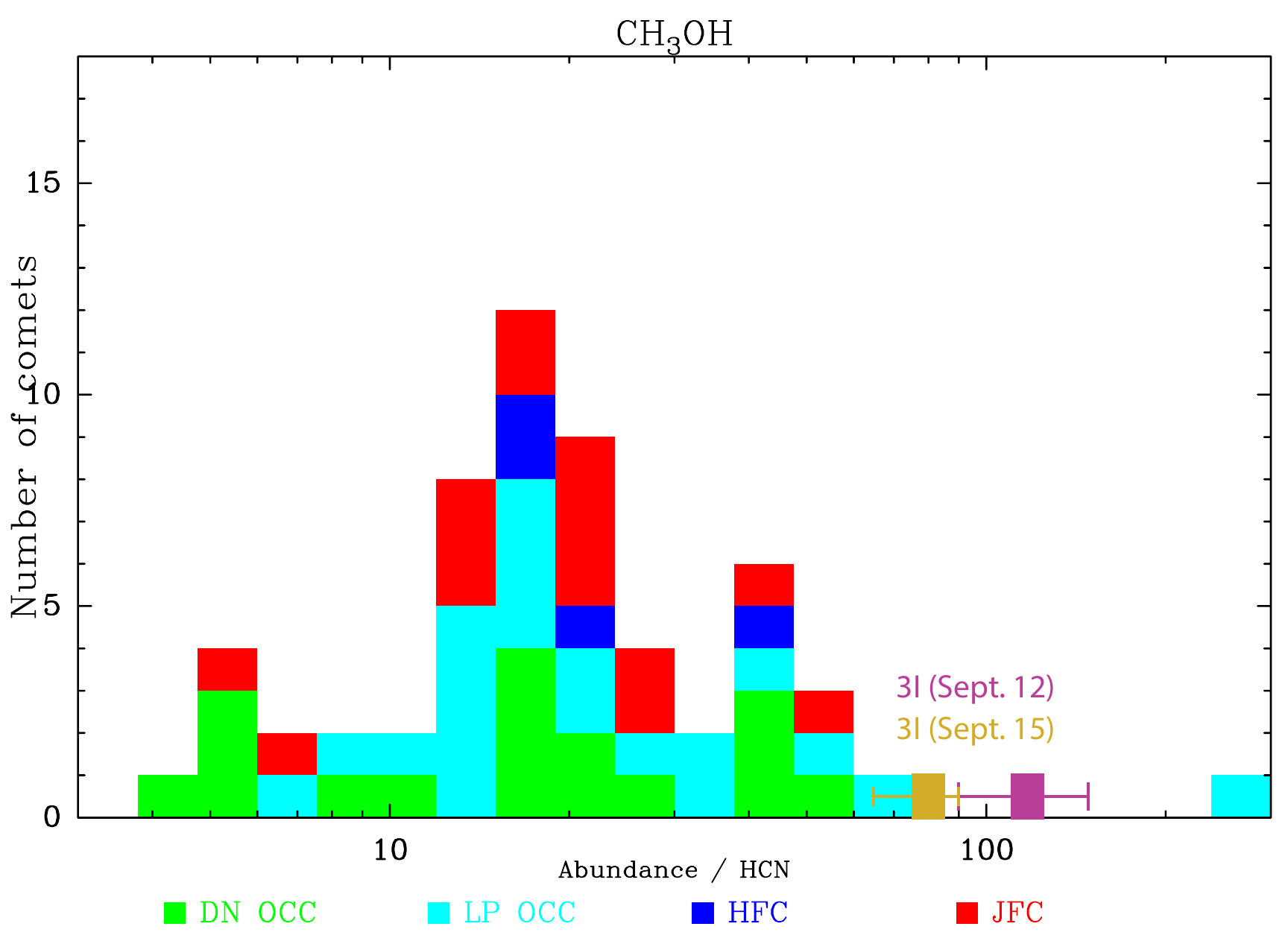}{0.45\textwidth}{(D)}
}
\caption{\textbf{(A).} HCN (\Ju{} = $4-3$) integrated flux map on September 12 with contours for the blue and red components of the line overlaid separately. The extracted spectral line profile shows the integration region for the blue and red maps across the line. Contours are in $3\sigma$ increments, with $3\sigma$ being the lowest contour level. \textbf{(B).} As in panel (A) for September 15. \textbf{(C).} $Q$(CH$_3$OH) in 3I/ATLAS as a function of \rh{} measured with the ACA. Also shown in the best-fit power law curve. \textbf{(D).} CH$_3$OH/HCN ratio in solar system comets measured at radio wavelengths to date adapted from \cite{Biver2024}. The values for 3I (CH$_3$OH/HCN=$124^{+30}_{-34}$ and $79^{+11}_{-14}$) are indicated with purple and yellow boxes for September 12 and 15, respectively, with the horizontal dashes indicating the $1\sigma$ uncertainty. The most enriched comet (far right) is C/2016 R2 (PanSTARRS), with CH$_3$OH/HCN=$280\pm72$ \citep{Biver2018}.
\label{fig:trend}}
\end{figure*}

Examining the trend in $Q$(CH$_3$OH) with \rh{} in 3I (Table~\ref{tab:kinematics}) shows a sharp increase from August 28 through October 1, including a significant uptick at the inner edge of the H$_2$O sublimation zone near \rh{} = 2 au, which may have coincided with changes in outgassing as H$_2$O activity in 3I became fully activated. Interestingly, \cite{Zhang2025} showed a rapid increase in coma brightness near \rh{} = 2 au based on monitoring with space-based solar observatories. Figure~\ref{fig:trend}C shows the evolution of $Q$(CH$_3$OH) in 3I/ATLAS, which is best fit by an \rh{}$^{-5.2\pm0.6}$ dependence, consistent with a dramatic rise in its coma methanol content as it approached perihelion.

In terms of the CH$_3$OH/HCN ratio in 3I, it is interesting to compare against values measured in solar system comets. Both molecules are commonly measured at radio and infrared wavelengths in solar system comets. Although there is general agreement for an average CH$_3$OH/H$_2$O solar system value of $\sim2$\% between both wavelengths, there is often significant disagreement on the HCN/H$_2$O abundance in a given comet between IR and radio observations (up to a factor of 6), with the IR population mean being $\sim0.2$\% and the radio mean being $\sim0.1$\% \citep{Biver2024}. Using our derived $Q$(CH$_3$OH)=$(378\pm116)\times10^{26}$\rh{}$^{-5.2\pm0.6}$ \ps{} relationship (Figure~\ref{fig:trend}C), we calculated $Q$(CH$_3$OH)=$(6.6^{+0.4}_{-0.9})\times10^{26}$ \ps{} and $(8.2^{+0.3}_{-1.0})\times10^{26}$ \ps{} on September 12 and 15, respectively. Using $Q$(HCN) from the best-fit 1D models on these dates, this gives CH$_3$OH/HCN=$124^{+30}_{-34}$ and $79^{+11}_{-14}$. 

Figure~\ref{fig:trend}D shows that 3I was highly enriched in its CH$_3$OH/HCN abundance relative to solar system comets measured in the radio. Our CH$_3$OH/HCN abundance for 3I is $1.2-4.7\sigma$ above the solar system comet mean value of $26\pm27$. However, it is important to note that our value for 3I was measured at \rh{} = 2.17 and 2.08 au when 3I may have still been transitioning from CO$_2$- to H$_2$O-dominated outgassing \citep{Cordiner2025a}. In contrast, most solar system comets are observed at smaller \rh{} with a fully H$_2$O-dominated coma, and comparisons between comets at larger \rh{} are not necessarily straightforward. However, \cite{Minissale2022} showed that CH$_3$OH is less volatile than HCN on H$_2$O-ice surfaces, and it is plausible to expect that CH$_3$OH/HCN may generally increase in a given comet with decreasing \rh{}. Although the nominal values for 3I on September 12 and 15 show the opposite trend, they are indistinguishable from each other within $2\sigma$ uncertainty and the difference in \rh{} between the two dates is small. Interestingly, the most highly enriched comet was C/2016 R2 (PanSTARRS), which showed a highly anomalous composition compared to other solar system comets and was measured at a somewhat larger \rh{} = 2.83 au \citep{Biver2018} than our observations of 3I. 

We can also use our $Q$(CH$_3$OH)(\rh{}) relationship to calculate CH$_3$OH/H$_2$O = $(8\pm2)\%$ at the \rh{} = 1.43 au of the OH 18 cm observations by \cite{Crovisier2025}, affirming that 3I is a CH$_3$OH-enriched object compared to solar system comets while keeping in mind the caveats of comparing non-simultaneous measurements.

\section{Conclusion} \label{sec:conclusion}
Our study of 3I/ATLAS from \rh{} = 2.6 au to 1.7 au pre-perihelion revealed how its CH$_3$OH production rate changed with time, and placed it into context with solar system comets. We found that HCN and CH$_3$OH displayed distinct outgassing kinematics, which may be related to a heterogeneous nucleus or to projection effects. At 99\% confidence, HCN production was indistinguishable from direct nucleus sublimation, whereas CH$_3$OH had contributions from a coma source at $L_p>258$ km from the nucleus; however, solely direct nucleus release of CH$_3$OH cannot be completely ruled out. The CH$_3$OH production rate underwent a dramatic increase with \rh{} as 3I crossed the H$_2$O sublimation zone. Its CH$_3$OH/HCN ratios of $124^{+30}_{-34}$ and $79^{+11}_{-14}$ measured at \rh{} = 2.17 au and 2.08 au, respectively, are among the highest observed in a comet to date. Observing campaigns targeting the comet near perihelion and as its angular distance from the Sun increases post-perihelion will doubtless provide further clues into the composition of this visitor from another star.

%% file: obsAppendix.tex
\section{Observing Log, Calibration, and Imaging}\label{sec:obslog}

\begin{deluxetable*}{cccccccccccc}[h]
\tablenum{A1}
\tablecaption{Observing Log\label{tab:obslog}}
\tablewidth{0pt}
\tablehead{
\colhead{Date} & \colhead{UT Time} & \colhead{\rh{}} & \colhead{$\Delta$} & \colhead{$\phi_\mathrm{STO}$} & \colhead{$\psi_{\sun}$} & \colhead{Target} & 
\colhead{\textit{N}\subs{ants}}  & \colhead{Baselines} & \colhead{PWV} & \colhead{$\theta$\subs{min}} & \colhead{$\theta$\subs{min}} \\
\colhead{(2025)} & \colhead{} & \colhead{(au)} & \colhead{(au)} & \colhead{($\degr$)} & \colhead{($\degr$)} & \colhead{Species}  & \colhead{} & \colhead{(m)}  & \colhead{(mm)} & \colhead{($\arcsec$)} & \colhead{(km)}
}
\startdata
\hline
28 Aug  & 20:59--22:35 & 2.60 & 2.58 & 22.4 & 102.6 & CH$_3$OH & 8 & 8--49 & 0.56 & $4\farcs10\times3\farcs28$ & $7672\times6137$ \\ 
12 Sept & 18:22--19:56 & 2.17 & 2.54 & 22.9 & 103.6 & HCN$^*$ & 8 & 8--49 & 0.81 & $3\farcs67\times2\farcs85$ & $6638\times5135$ \\
15 Sept & 18:35--20:10 & 2.08 & 2.54 & 22.5 & 103.7 & HCN$^*$ & 9 & 8--49 & 1.35 & $4\farcs24\times2\farcs82$ & $7670\times5101$ \\
18 Sept & 17:07--18:43 & 2.01 & 2.53 & 21.8 & 103.7 & CH$_3$OH & 8 & 8--49 & 0.58 & $4\farcs19\times2\farcs80$ & $7688\times5137$ \\
18 Sept & 19:27--21:02 & 2.00 & 2.53 & 21.8 & 103.7 & CH$_3$OH$^*$ & 8 & 8--49 & 0.60 & $4\farcs30\times2\farcs64$ & $7890\times4844$ \\
18 Sept & 21:03--22:36 & 2.00 & 2.53 & 21.8 & 103.7 & CH$_3$OH$^*$ & 8 & 8--49 & 0.61 & $5\farcs21\times2\farcs62$ & $9560\times4807$ \\
22 Sept & 16:34--18:13 & 1.90 & 2.52 & 20.7 & 103.7 & CH$_3$OH & 9 & 8--49 & 0.69 & $4\farcs65\times3\farcs27$ & $8345\times5868$ \\
01 Oct & 19:29--21:09 & 1.69 & 2.49 & 16.6 & 102.8 & CH$_3$OH & 9 & 8--49 & 0.69 & $5\farcs26\times3\farcs09$ & $9327\times5479$ 
\enddata
\tablecomments{\rh{}, $\Delta$, $\phi_\mathrm{STO}$, and $\psi_{\sun}$ are the heliocentric distance, geocentric distance,
phase angle (Sun--Comet--Earth), and position angle of the Sun--Comet radius vector, respectively, of 3I at the time of observations. The targeted molecule (HCN or CH$_3$OH) is indicated for each execution, and $^*$ denotes those using high (122 kHz channel spacing) spectral resolution. \textit{N}\subs{ants}
is the number of antennas utilized during each observation, with the range of baseline lengths indicated for each. PWV is the mean precipitable water vapor at zenith during the observations. $\theta$\subs{min} is the angular resolution (synthesized beam) at $\nu$ given in arcseconds and in projected distance (km) at the geocentric distance of 3I. The total on-source integration time was 49 minutes for all epochs.}
\end{deluxetable*}

Table~\ref{tab:obslog} summarizes observing circumstances for the ACA observations of 3I/ATLAS. Observations on August 28, September 18 and 22, and October 1 used four 1875 MHz wide spectral windows with a frequency resolution of 488 kHz. These executions simultaneously sampled multiple transitions of the CH$_3$OH ($J_K=7_K-6_K$) ladder along with the ($J_K=1_1-0_0 A^+$) and ($J_K=4_0-3_{-1} E$) transitions (Table~\ref{tab:lines}), thereby enabling a test of the coma kinetic temperature. Observations on September 12 and 15 targeted the HCN (\Ju{}=4--3) transition, and additional observations on September 18 measured the CH$_3$OH ($J_K=1_1-0_0 A^+$) transition, each using one 244 MHz wide spectral window with a spectral resolution of 122 kHz. A channel averaging factor ($N$) of 2 was applied to the HCN spectra at the correlator (resulting in a spectral resolution of 0.13 \kms{}), whereas a channel averaging factor $N=1$ was used for CH$_3$OH \citep[resulting in a spectral resolution of 0.21 \kms{} for the high resolution spectra and 0.87 \kms{} for the low resolution spectra. See Chapter 5 of the ALMA Technnical Handbook;][]{Cortes2024}. We tracked the comet position using JPL Horizons ephemerides (JPL \#25 for August 28 and \#26 for September dates). An observing log is shown in Table~\ref{tab:obslog}.

The mean precipitable water vapor at zenith (zenith PWV) ranged from 0.56--1.35 mm. Quasar observations were used for bandpass and phase calibration for all epochs.  Asteroid Vesta was used to calibrate the flux scale for the low spectral resolution CH$_3$OH observations. Quasars (J1337-1257 for both executions of the CH$_3$OH setting and J1256-0547 for both executions of the HCN setting) were used to calibrate the flux scale for the high spectral resolution observations. We examined the ALMA Calibrator Catalog for both quasars and found no evidence for significant variability in their Band 7 (345 GHz) fluxes during measurements taken within several days of our observations. Furthermore, the same quasar was used for both executions of a given instrumental setting, minimizing uncertainties in the calibration from date to date. The spatial scale (the range in semi-minor and semi-major axes of the synthesized beam) was 2$\farcs$80 -- 5$\farcs$26.

The data were flagged, calibrated, and imaged using standard routines in Common Astronomy Software Applications (CASA) package version 6.4.1 \citep{CASA2022}. We deconvolved the point-spread function with the Högbom algorithm, using natural visibility weighting, a 10$\arcsec$ diameter mask centered on the peak continuum position, and a flux threshold of twice the rms noise. The deconvolved images were then convolved with the synthesized beam and corrected for the (Gaussian) response of the ACA primary beam (FWHM $29\farcs3$ at 338 GHz and $27\farcs9$ at 354 GHz, respectively). We transformed the images from astrometric coordinates to projected cometocentric distances, with the location of the peak gas flux chosen as the origin, which was in good agreement with the comet's predicted ephemeris position ($<0\farcs75$ offset between the peak position and the array pointing phasecenter). 

%% file: mapsAppendix.tex
\section{Additional CH$_3$OH Maps}\label{sec:extraMaps}
\begin{figure*}
\gridline{\fig{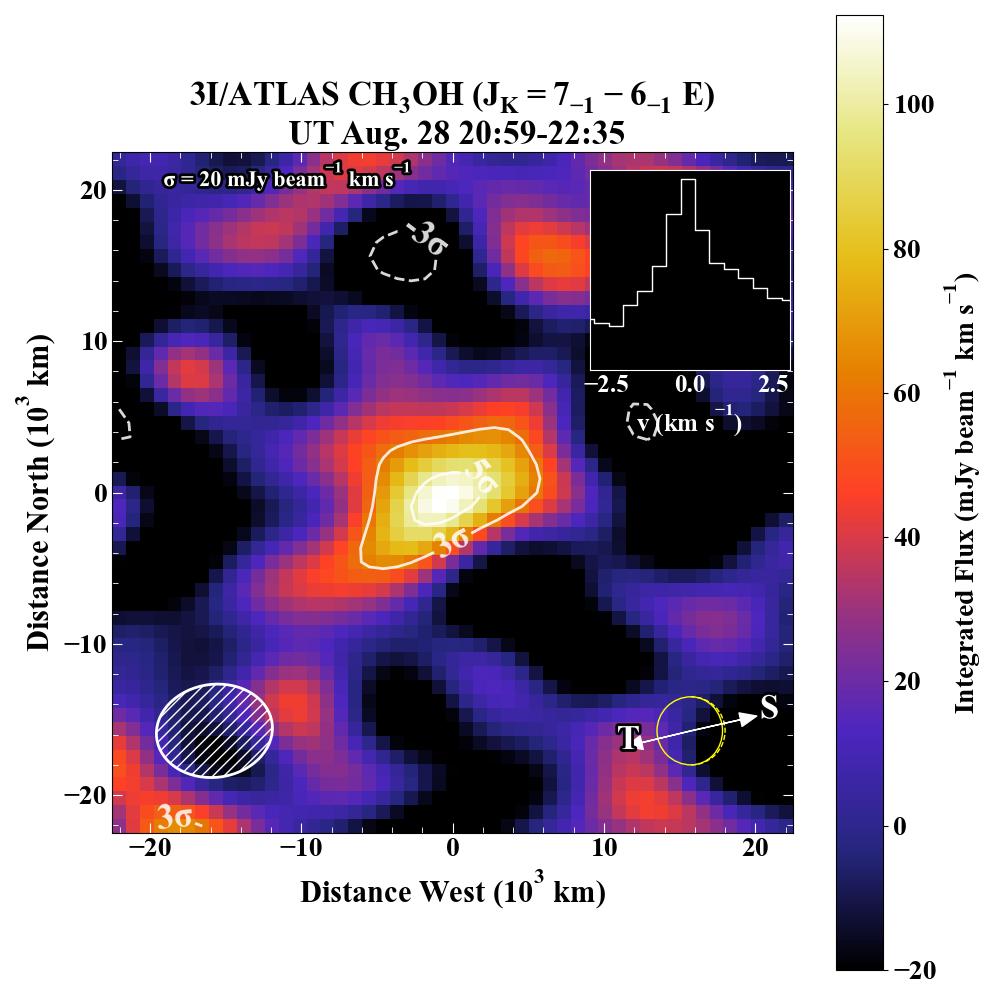}{0.45\textwidth}{(A)}
          \fig{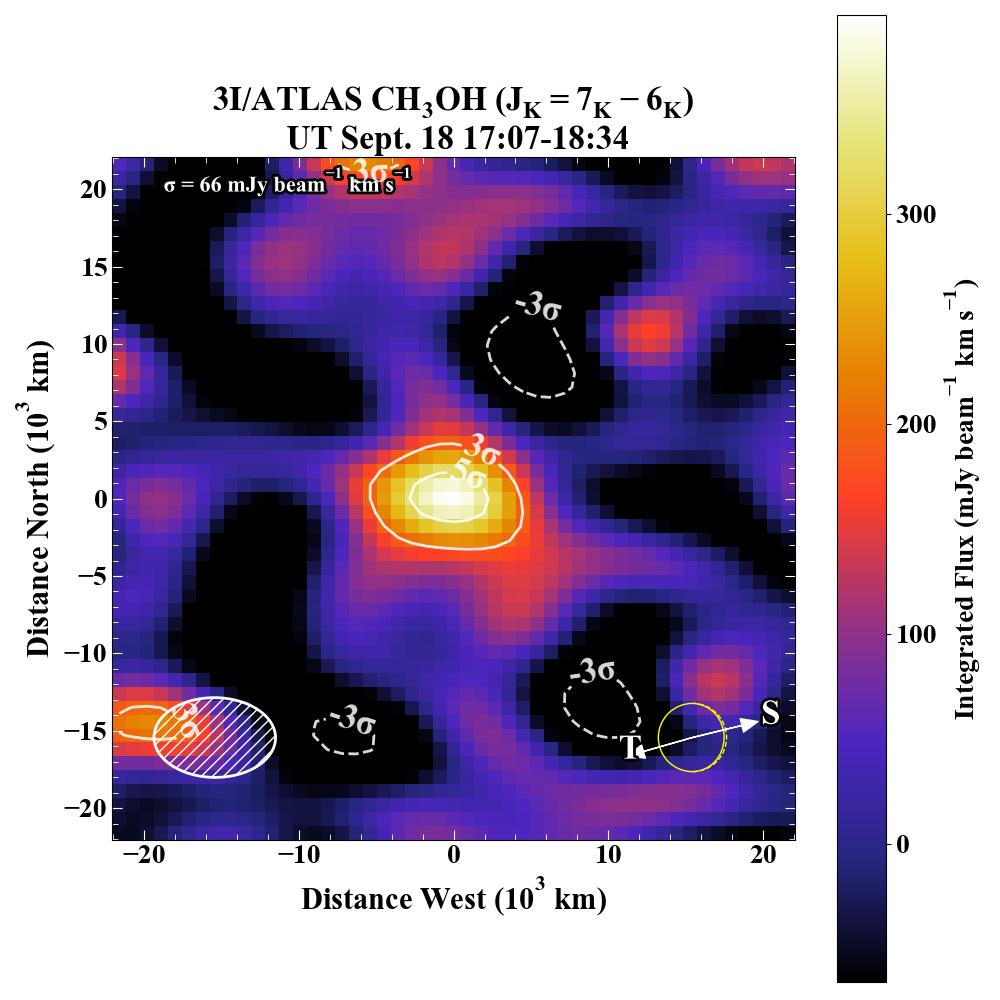}{0.45\textwidth}{(B)}
}
\gridline{\fig{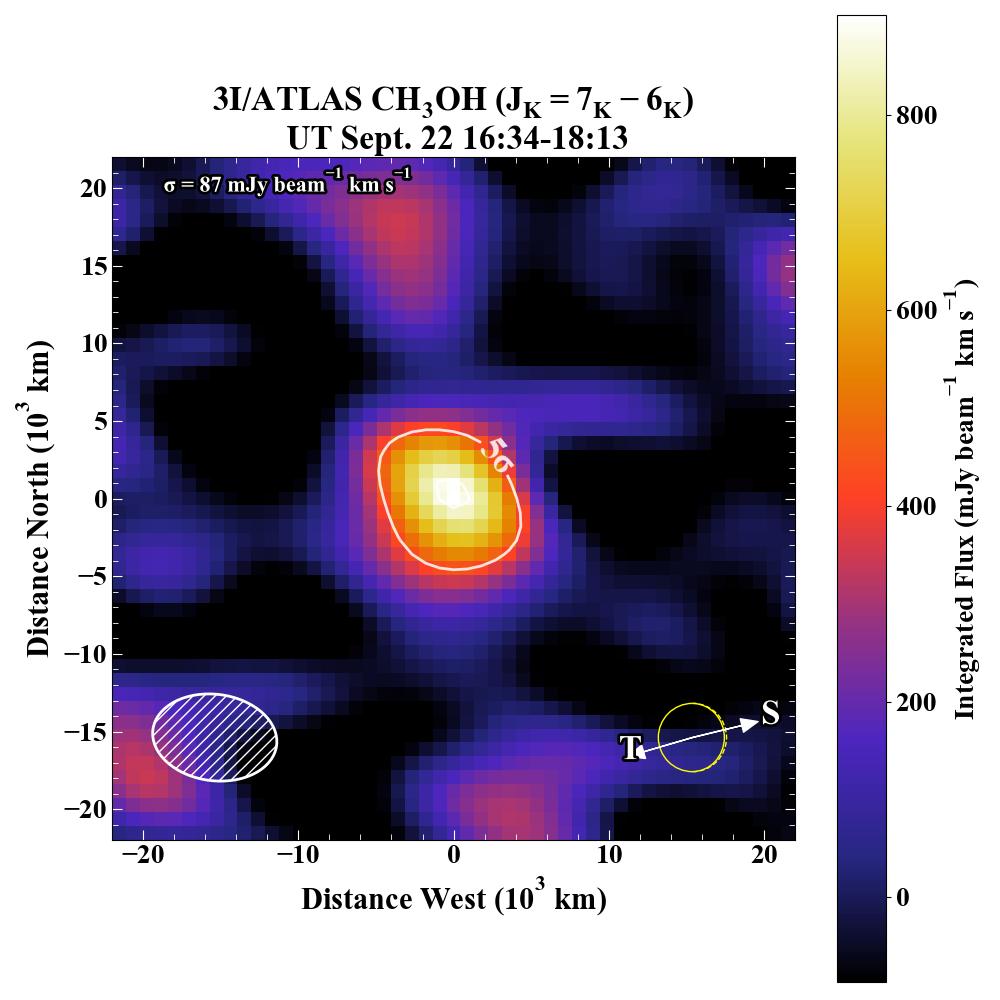}{0.45\textwidth}{(C)}
          \fig{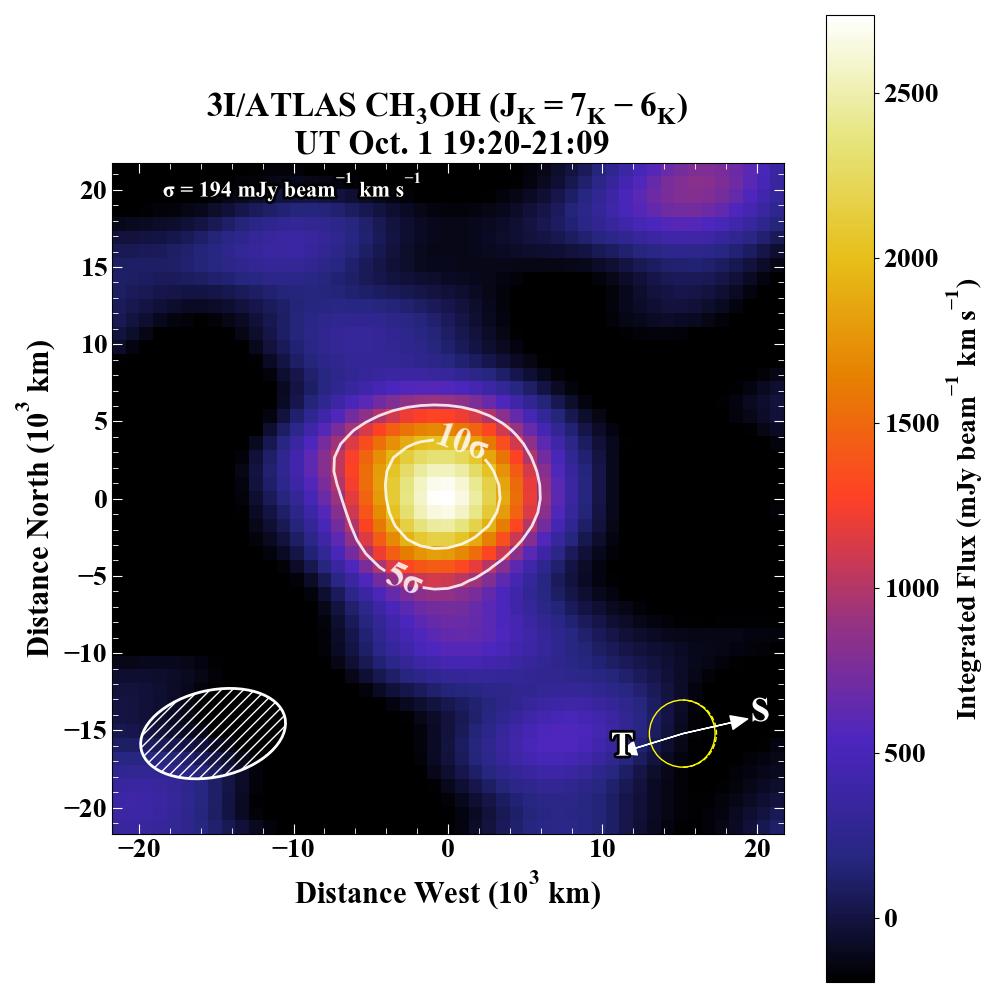}{0.45\textwidth}{(D)}
}
\caption{\textbf{(A)--(D).} Spectrally integrated flux maps for CH$_3$OH on August 28, September 18 and 22, and October 1, with traces and labels as in Figure~\ref{fig:main-maps}. Panel A shows a spectrum extracted in a $10\arcsec$ diameter aperture centered on the peak gas emission for the transition indicated. Panels B, C, and D show the integrated flux for all detected CH$_3$OH ($J_K=7_K-6_K$) transitions on a given date.
\label{fig:ch3oh-maps2}}
\end{figure*}

Figure~\ref{fig:ch3oh-maps2} provides additional maps of CH$_3$OH emission measured in 3I, and Table~\ref{tab:lines} details all detected transitions.

\begin{deluxetable*}{ccccc}
\tablenum{B1}
\tablecaption{Lines Detected in 3I/ATLAS \label{tab:lines}}
\tablewidth{0pt}
\tablehead{
\colhead{Species} & \colhead{Transition$^{(a)}$} & \colhead{Frequency} & \colhead{\textit{E}\subs{u}} & \colhead{Date}\\
\colhead{} & \colhead{} & \colhead{(GHz)} & \colhead{(K)} & \colhead{(2025)}
}
\startdata
\hline
HCN & $4-3$ & 354.505478 & 42.5 & Sept. 12, 15\\
CH$_3$OH & $1_1-0_0 A^+$ & 350.905100 & 16.8  & Sept. 18, 22, Oct. 1 \\
         & $4_0-3_{-1} E$ & 350.687662 & 36.3 & Sept. 18, 22, Oct. 1  \\
         & $7_0-6_0 A^+$ & 338.408698 & 64.9 & Sept. 18, 22, Oct. 1 \\
         & $7_{-1}-6_{-1} E$ & 338.344588 & 70.6 & Aug. 28, Sept. 18, 22, Oct. 1 \\
         & $7_0-6_0 E$ & 338.124488 & 78.1 & Sept. 18, 22, Oct. 1 \\
         & $7_1-6_1 E$ & 338.614936 & 86.1 &  Oct. 1 \\
         & $7_2-6_2 E$ & 338.721693 & 87.3 & Sept. 18, 22, Oct. 1 \\
         & $7_{-2}-6_{-2} E$ & 338.722898 & 90.9 & Sept. 18, 22, Oct. 1 \\
         & $7_2-6_2 A^-$ & 338.512853 & 102.7 & Oct. 1 \\
         & $7_2-6_2 A^+$ & 338.639802 & 102.7 & Oct. 1 \\
         & $7_3-6_3 E$ & 338.583216 & 112.7 & Oct. 1 \\
         & $7_3-6_3 A^+$ & 338.540826 & 114.8 & Oct. 1 \\
         & $7_3-6_3 A^-$ & 338.543152 & 114.8 & Oct. 1 \\
\enddata
\tablecomments{\sups{a} Quantum numbers are given as ($J'-J''$) for HCN and as ($J'_{K'}-J''_{K''}$) for CH$_3$OH.
}
\end{deluxetable*}

%% file: radAppendix.tex
\section{Fourier Domain Modeling}\label{sec:fourier}
We chose to perform modeling of the interferometric spectra in the Fourier domain to avoid the introduction of imaging artifacts arising from incomplete $uv$ sampling inherent to interferometric observations and the CLEAN algorithm itself \citep{Boissier2007,Bockelee2010,Cordiner2025b}. Our procedure, which we outline briefly here, is described in detail in \cite{Cordiner2023}. We first time-averaged the observed visibilities. We then used the \texttt{vis\_sample} program \citep{Loomis2018} to take the Fourier transform of our radiative transfer model cubes using the same $uv$ coverage as our (time averaged) ALMA observations (i.e., model fitting was performed to the time-averaged $uv$ sampling without any averaging over baseline length). We then performed least-squares fits of our radiative transfer models against the measured interferometric visibilities. We used the \texttt{lmfit} application of the Levenberg-Marquardt minimization technique and retrieved uncertainties on our optimized parameters from the diagonal elements of the covariance matrix, minimizing the residual for the real and imaginary parts of the observed and modeled interferometric visibilities. We normalized the residuals by the measured per-baseline noise at each iteration of the fitting process.

\subsection{Determination of CH$_3$OH Outgassing Geometry}\label{subsec:methanol-outgas}
For CH$_3$OH, we tested the application of both 1D and 3D modeling of the high spectral resolution data.  For the 3D models, following the formalism described in \S~\ref{subsec:radiative}, we divided the coma into two regions, $R_1$ and $R_2$. Region $R_1$ is defined as the conical region with half-opening angle $\gamma$, originating at the nucleus surface and oriented at a phase angle $\phi$ with respect to the observer and position angle $\psi$ in the plane of the sky \citep[see Fig. 8 of][]{Cordiner2023}. We assumed hemispheric asymmetry along the projected Sun-comet vector ($\phi=\phi_\mathrm{STO}$, $\psi=\psi_{\sun}$, $\gamma=90\degr$). We assumed a gas kinetic temperature $T_{\mathrm{kin}}$ = 60 K as a first pass based on inner coma (nucleocentric distances $<$5900 km) rotational temperatures reported in \cite{Cordiner2025a}. We determined the CH$_3$OH distribution in $R_1, R_2$ using a Haser formalism \citep{Haser1957}:
\begin{equation}
    n_d(r) = \frac{Q_i}{4 \pi v_i r^2}\frac{\frac{v_i}{\beta_d}}{\frac{v_i}{\beta_p}-\frac{v_i}{\beta_d}}\left[\exp{\left(-\frac{\beta_p}{v_i}r\right)}-\exp{\left(-\frac{\beta_d}{v_i}r\right)}\right],
\end{equation}
\noindent where $Q_i$ and $v_i$ are the molecular production rate (s$^{-1}$) and gas expansion velocity (km\,s$^{-1}$) in each coma region, and $\beta_p, \beta_d$ are the parent and daughter molecular photodissociation rates (s$^{-1}$), respectively. 

We first worked to constrain the kinematics, assuming a parent model and letting the molecular production rate, the ratio $Q_1/Q_2$ of production rates between $R_1$ and $R_2$, $v_1,$ and $v_2$ vary as free parameters. We found $Q_1/Q_2=1.6\pm0.5$, $v_1=0.62\pm0.06$ \kms{}, $v_2=0.33\pm0.04$ \kms{}. Similarly for a 1D parent model, we let the production rate, $v$, and $\Delta v$ vary as free parameters. We found $v=0.48\pm0.03$ \kms{} and $\Delta v=-0.13\pm0.03$ \kms{}. 

We then worked to determine the kinetic temperature, leveraging the higher number of lines $(J_K=7_K-6_K$ ladder near 338 GHz) sampled by the low-resolution spectra measured on the same date. We took $v$ and $\Delta v$ found from the 1D model of the high resolution spectra and fixed those values while running a 1D parent model of the low resolution spectra. Owing to a lack of signal on long baselines, we assumed an isothermal profile and let the production rate and kinetic temperature vary as free parameters, finding \tkin{}=$49\pm12$ K.

\subsection{Determination of CH$_3$OH Production Mechanisms}\label{subsec:ch3oh-parents}
We then returned to the high resolution spectra to test for coma production of CH$_3$OH while adopting \tkin{} = 49 K. The density of coma parent molecules peaks at the nucleus before falling off exponentially with increasing nucleocentric distance owing to adiabatic expansion and photolysis. To adequately sample the often asymmetric (and nonlinear) uncertainties on $L_p$ we followed the methods of \cite{Cordiner2023}, generating the $\chi^2$ surface for the 1D and 3D models as a function of $L_p$. We iterated over a range of $L_p$ values, performing least-squares fitting where the production rate and expansion speed were allowed to vary while holding all other parameters fixed. We recorded the $\chi^2$ statistic of the optimal SUBLIME models (calculated as the goodness of fit for the observed vs.\ model visibilities) found for each fixed $L_p$ value. We then examined the $\Delta\chi^2$ curve generated using cubic spline interpolation between each value of $L_p$, where $\Delta\chi^2(L_p) = \chi^2(L_p) - \chi^2_{min}$. This procedure is described in \cite{Avni1976}, where the increment in $\chi^2$ about $\chi^2_{min}$ corresponding to a given confidence interval depends only on the number of parameters being simultaneously estimated (not the number of parameters in the fitting function) and is independent of the value of $\chi^2_{min}$ itself. In our case, the value of $L_p$ is the sole parameter being estimated, so we obtained the 1$\sigma$ and 2.6$\sigma$ uncertainties from the values for $\Delta\chi^2$ = 1 (68\% confidence) and 6.63 (99\% confidence) thresholds. Our $\Delta\chi^2$ curves are shown in Figure~\ref{fig:ch3oh-models}, and indicate that at 99\% confidence there was a source of coma production of CH$_3$OH at $L_p>258$ km. The best-fit $Q$(CH$_3$OH) and $L_p$ are in good agreement between the 1D and 3D models and are tabulated in Table~\ref{tab:kinematics}. 

Finally, we worked to determine whether the 3D model provided a statistically improved fit to the CH$_3$OH emission over the 1D model. From a statistical standpoint, these may be viewed as a case of nested models: the 3D model reduces to the 1D model in the case that $Q_1/Q_2=1$ and $v_2=v_1$. An F-test is an appropriate comparison of the relative goodness of fit for nested models, where the null hypothesis is that the simpler (1D) model provided a better fit to the data. We performed an F-test calculating the residual sum of squares (RSS) between each model and the data and determining the F value as $F = \frac{RSS_{\mathrm{1D}}-RSS_{\mathrm{3D}}}{RSS_{\mathrm{3D}}}\frac{n-p_{\mathrm{3D}}}{p_{\mathrm{3D}}-p_{\mathrm{1D}}}$, where $n$ is the number of visibilities and $(p_{\mathrm{3D}},p_{\mathrm{1D}})$ are the number of parameters in the 3D and 1D models, respectively. For $\alpha=0.05$ (95\% confidence), the critical value for an F-distribution with $(p_{\mathrm{3D}}-p_{\mathrm{1D}},n-p_{\mathrm{3D}})$ degrees of freedom is 2.99, whereas our F-value is 1.45. Thus, we cannot reject the null hypothesis. Nevertheless, both models give a consistent overall picture of the CH$_3$OH outgassing geometry in 3I. A model-data comparison of the visibility amplitudes as a function of baseline for the 1D parent and best-fit daughter models is given in Figure~\ref{fig:ch3oh-vis}.

\subsection{Application to Low Resolution CH$_3$OH Spectra}\label{subsec:lowres}
We fixed the best-fit $v, \Delta v$, and $L_p$ from the 1D model of the September 18 high-resolution $J_K=1_1-0_0 A^+$ spectra when running a 1D daughter model of the low-resolution $J_K=7_K-6_K$ spectra on the same date, letting the production rate vary as a free parameter. Our resulting $Q$(CH$_3$OH) is in good agreement with that derived from the high-resolution spectra (Table~\ref{tab:kinematics}).

To model the remaining CH$_3$OH low resolution spectra on other dates, we used a 1D model and assumed that $v$ varied as \rh{}$^{-0.5}$ from the value retrieved from the high-resolution September 18 spectra, following trends measured in other comets \citep{Ootsubo2012} and allowing \tkin{} and $Q$(CH$_3$OH) to vary as free parameters. Unfortunately, the necessity to use an assumed expansion speed precludes a retrieval of $L_p$ on these dates due to its dependence on $v$. If the CH$_3$OH precursor were a gas-phase species, we could assume insolation-driven dependence (\rh{}$^2$) and scale $\beta_p$ to calculate an assumed $L_p$ on each date. However, the $L_p$ (and in turn, $\beta_p$) retrieved from the high resolution spectra on September 18 is inconsistent with that for any known cometary molecule with photolysis pathways to CH$_3$OH \citep{Hrodmarsson2023}, making it likely that the precursor was sublimation from icy grains (see \S~\ref{subsec:methanol-kinematics}). Since the velocities of the gas and dust decouple at small nucleocentric distances \citep{Bodewits2024}, we cannot calculate an assumed $L_p$ on August 28, September 22, and October 1. Thus, we used a parent model when applying 1D models to the low resolution CH$_3$OH spectra on all dates except September 18. Aside from the practical difficulties in scaling $L_p$ from September 18 to other dates, this is justified given that $L_p$ is considerably smaller than the size of the ACA synthesized beam (Table~\ref{tab:obslog}). The inclusion of a daughter source on these dates would result in a higher $Q$(CH$_3$OH) and would only reinforce our conclusion that 3I is highly enriched compared to solar system comets. Our best-fit models are shown in Figures~\ref{fig:lores-1}--\ref{fig:lores-6}.

\subsubsection{Determination of Kinetic Tempature}\label{subsubsec:tkin}
For August 28 (when only the $J_K=7_{-1}-6_{-1} E$ transition was detected) we assumed \tkin{} = 40 K based on our retrievals at smaller \rh{}. The non-detection of the $J_K=7_0-6_0 A^+$ transition, expected to be comparably strong, may be due to sub-optimal signal-to-noise and the low production rate during these observations at the largest \rh{} within the dataset. The coincidence of the emission with the expected ephemeris position of 3I, the overall similarity of the line profiles, the shape of the integrated flux, and the secure detection of the same line on multiple dates in September and October adds additional credibility.

\begin{figure}
\plotone{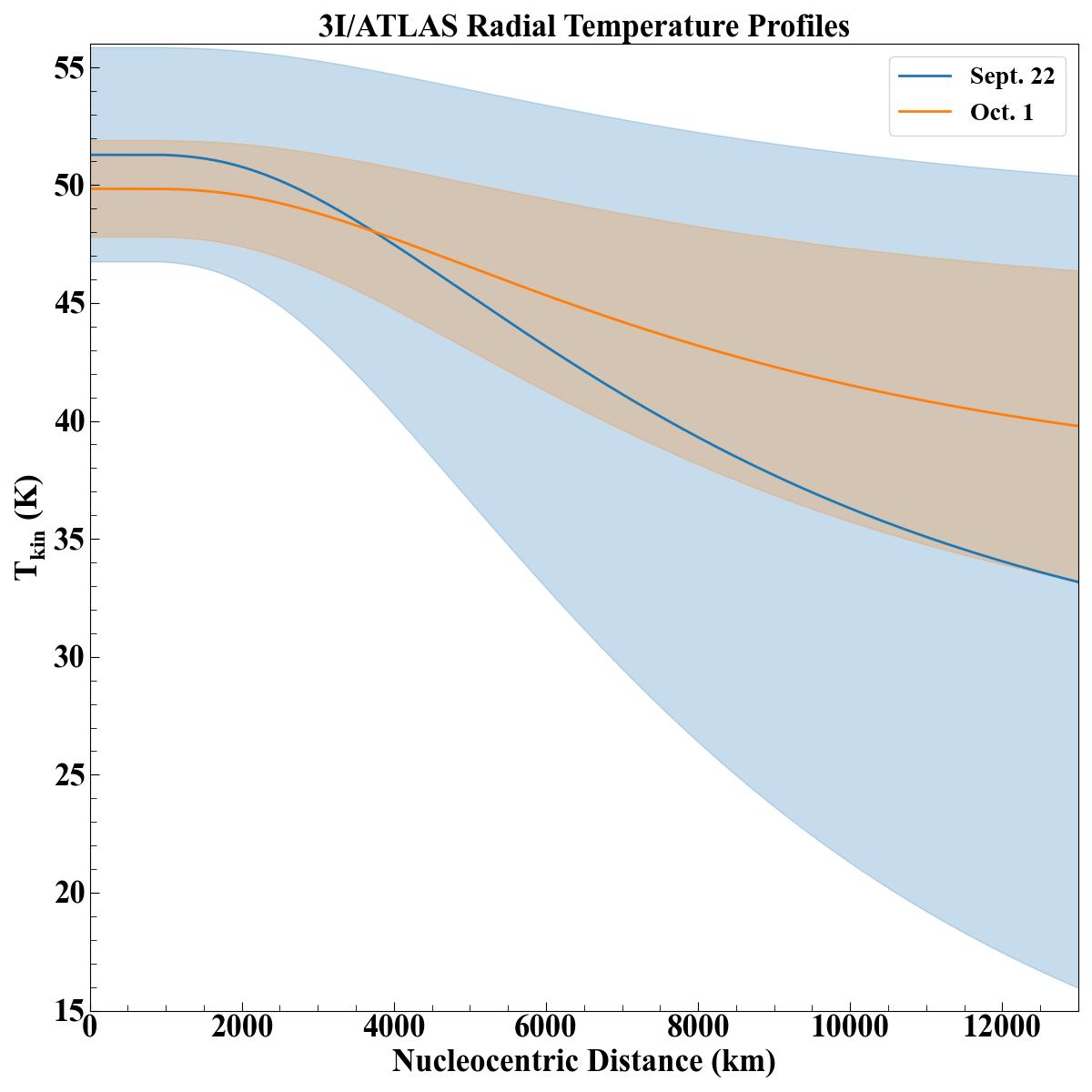}
\caption{Best-fit coma radial temperature profiles for CH$_3$OH in 3I on September 22 and October 1. The shaded regions represent $1\sigma$ uncertainties. 
\label{fig:coma-temp}}
\end{figure}

For September 18, we let \tkin{} vary as a free parameter assuming an isothermal profile given the lack of strong signal on the longer ACA baselines. We found \tkin{} = $49\pm12$ K. On September 22 and October 1, we tested the possibility of retrieving radial temperature profiles for CH$_3$OH given the increasing comet signal compared to earlier dates. Following \cite{Cordiner2023}, we used a smoothed, segmented, linear $T(r)$ function (in $\log(r)-T$ space where $r\in[\theta_\mathrm{min},\theta_\mathrm{MRS}])$ using a variable number of segments $(n)$ of equal length $(l_s)$ in $\log(r)$ space. Smoothing was performed (in $\log(r)$ space) with a Gaussian of FWHM equal to $l_s$. We incrementally increased the model complexity from the simplest assumption of a constant temperature until we obtained an acceptable fit to the data. This approach minimizes the number of model free parameters while ensuring sufficient degrees of freedom to reproduce the data. We avoided ill-conditioned models (with too many segments) by testing for large-amplitude ripples in $T(r)$. 

We set the temperature at a constant value interior to a radius equal to half the minor axis of the synthesized beam  $(r=\theta_\mathrm{min}/2)$ and exterior to a radius equal to half the maximum recoverable scale (MRS). A good fit was found  for $n=2$ radial points on both dates (being $r=[3500$ km, 12,870 km]). The best-fit temperature points are $T = [51\pm5$ K, $28\pm20$ K] and $T = [50\pm2$ K, $37\pm8$ K] on September 22 and October 1, respectively. Figure~\ref{fig:coma-temp} shows the best-fit two-segment temperature profile for CH$_3$OH in 3I on each date.  

\subsection{HCN Modeling}\label{subsec:hcn-models}
We followed a similar approach to modeling the high-resolution HCN spectra, applying both 3D and 1D SUBLIME models. With our nearest retrieved temperature being \tkin=$49\pm12$ K from CH$_3$OH on September 18 (\rh{} = 2.01 au), we assumed an isothermal \tkin{} = 45 K on September 12 and 15 (\rh{} = 2.17 au and 2.08 au, respectively) when modeling HCN. We then followed an identical formalism as for CH$_3$OH. For the 1D model, we determined a radial velocity offset, $\Delta v$, from application of a parent model, followed by a $\Delta\chi^2$ analysis to test for production of HCN from direct nucleus sublimation vs.\ coma sources. We did the same for the 3D model, except that no radial velocity offset was allowed. Our results are shown in Figure~\ref{fig:hires-models}. A comparison of the visibility amplitudes as a function of baseline for the parent and best-fit daughter 1D models on each date is given in Figure~\ref{fig:hcn-vis}. The best-fit 1D and 3D parent models for each date are shown in Figure~\ref{fig:hcn-compare}.

We also worked to test for evidence of coma gas acceleration as a mechanism to explain the underfitting of the HCN line flux on the longest baselines. We fit Gaussian profiles to the HCN line as a function of baseline on September 15 (owing to the higher S/N compared to September 12) and retrieved the line FWHM. Figure~\ref{fig:hcn-fwhm} shows that the line FWHM is consistent (within uncertainty) with $\sim0.4$ \kms{} out to $\theta<10\farcs5$ ($<19,00$ km; the exception being the low S/N spectra at 38 m baselines) before increasing significantly to $0.65\pm0.06$ \kms{} on the largest angular scales ($\theta<19\farcs8$). This is qualitatively consistent with the signatures of coma gas acceleration as seen in other comets \citep{Lammerzahl1987,Tseng2007,Cordiner2025b}.  Determining the $v(r)$ dependence for HCN in 3I will require additional observations and modeling and is beyond the scope of this work.

\begin{figure}
\plotone{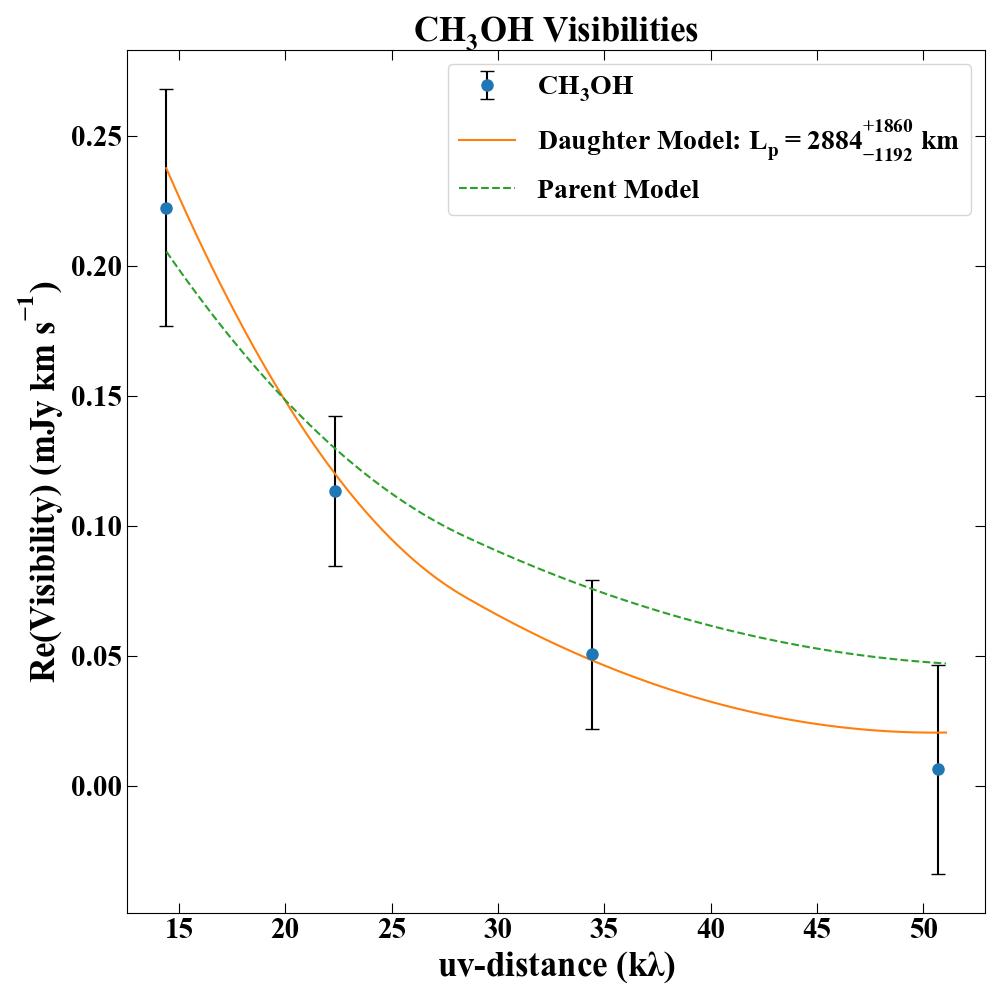}
\caption{CH$_3$OH $J_K=1_1-0_0 A^+$ visibilities as a function of $uv$-distance (expressed in $k\lambda$) and binned over the same baseline ranges as in Figure~\ref{fig:ch3oh-models}. The best-fit 1D daughter model and a parent model are shown.
\label{fig:ch3oh-vis}}
\end{figure}

\begin{figure}
\plotone{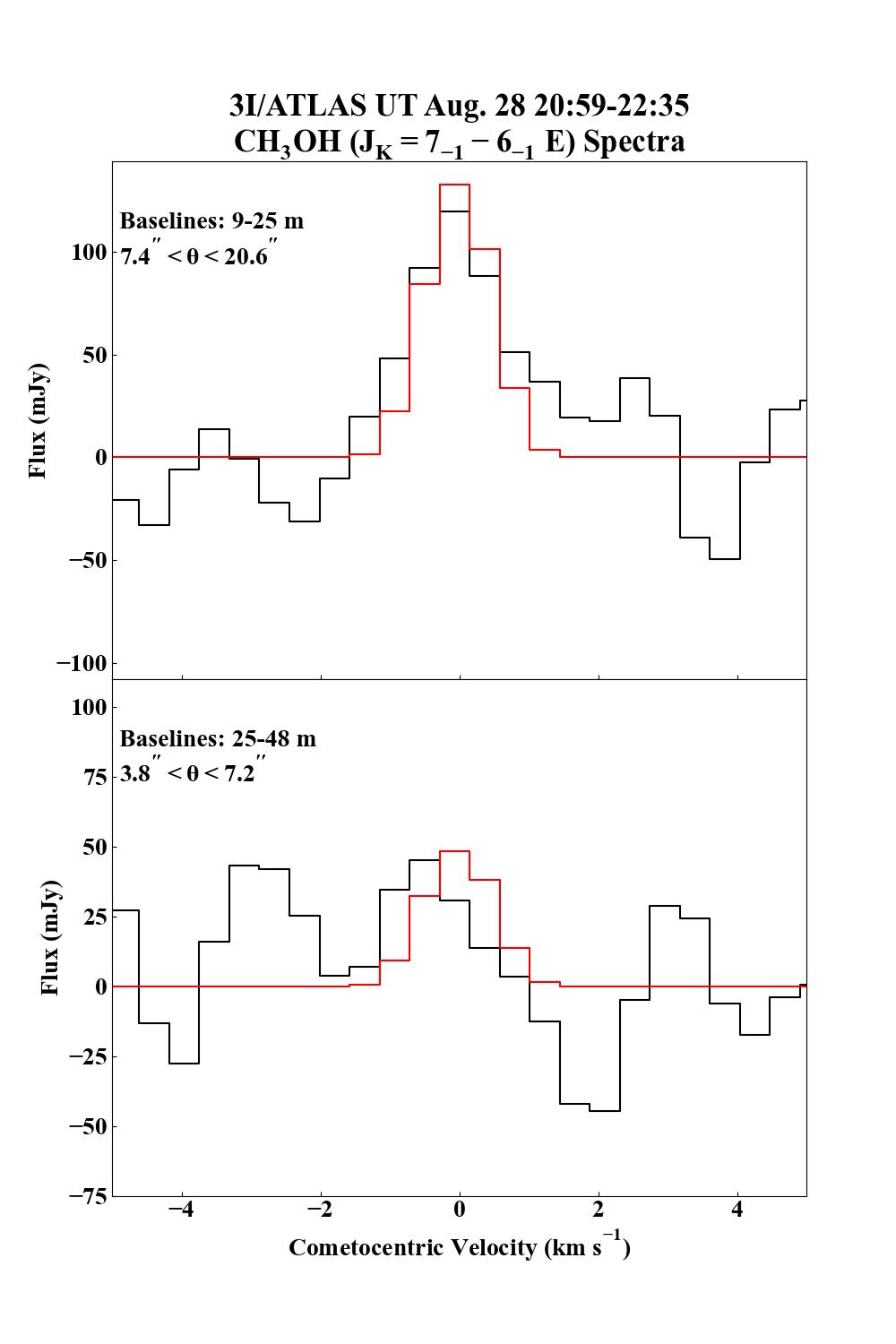}
\caption{CH$_3$OH spectrum in 3I on August 28, with each panel representing spectra extracted from differing baseline ranges (angular scales). The best-fit model is overlaid in red.
\label{fig:lores-1}}
\end{figure}

\begin{figure}
\plotone{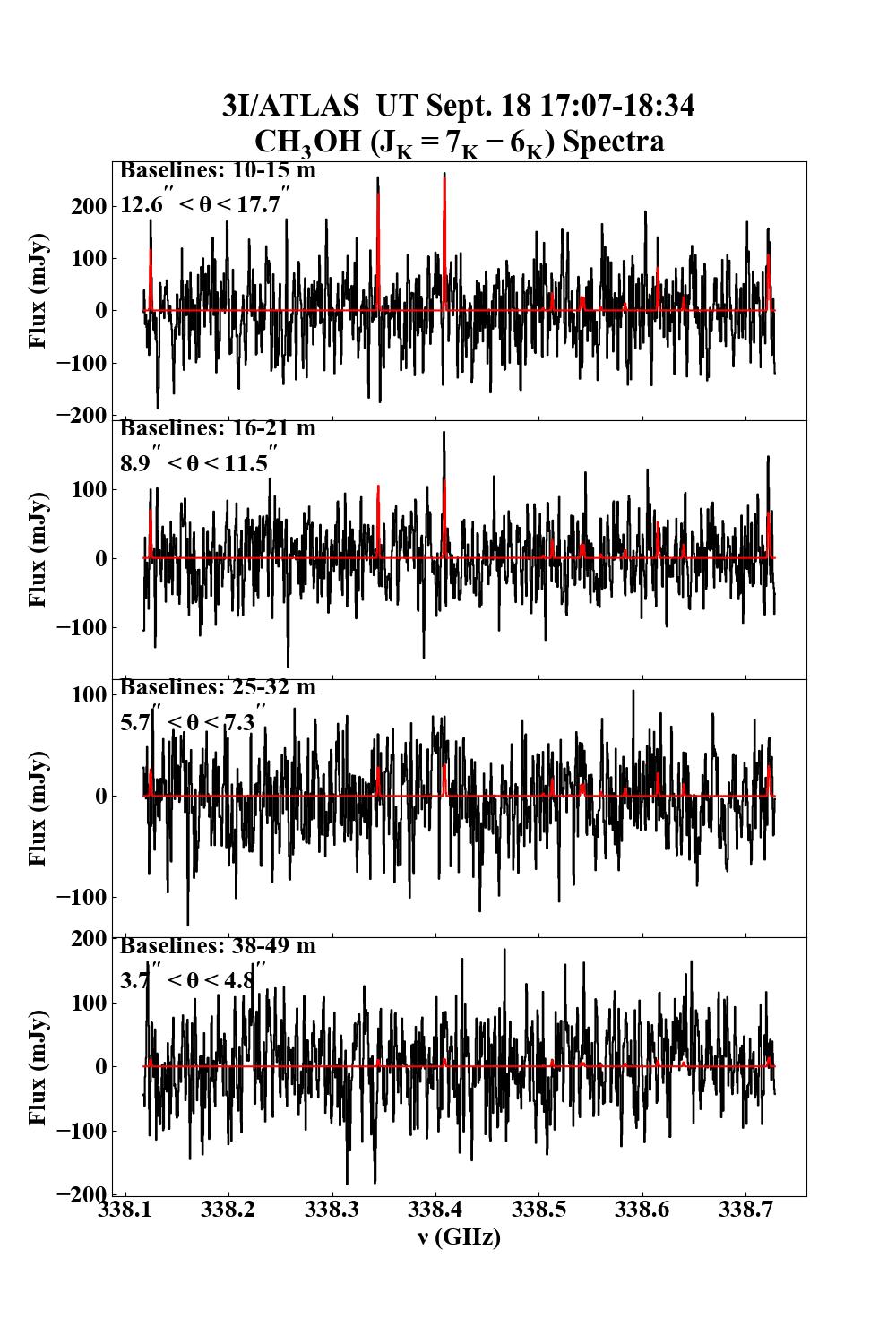}
\caption{CH$_3$OH spectrum in 3I on September 18, with each panel representing spectra extracted from differing baseline ranges (angular scales). The best-fit model is overlaid in red.
\label{fig:lores-2}}
\end{figure}

\begin{figure}
\plotone{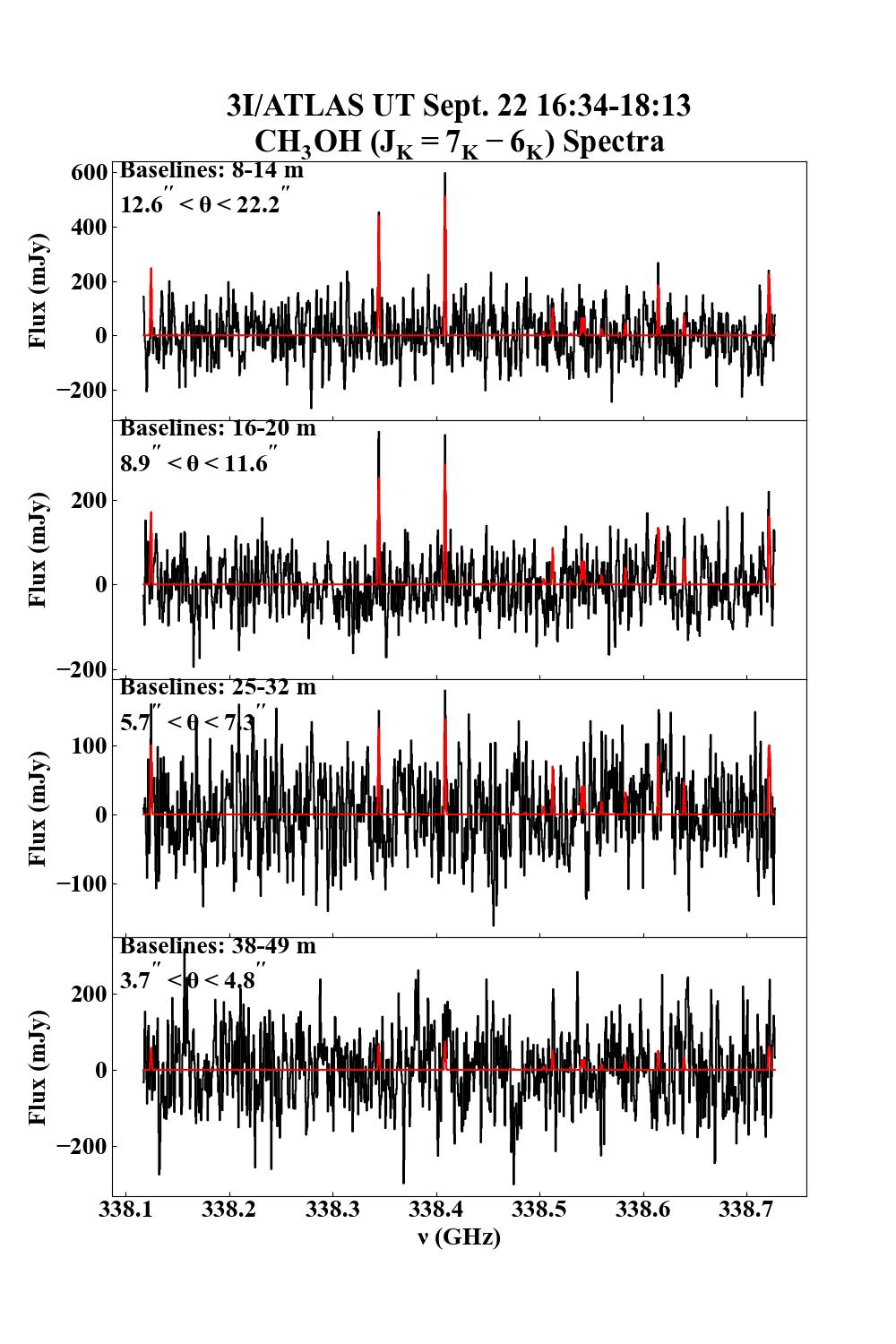}
\caption{CH$_3$OH spectrum in 3I on September 22, with each panel representing spectra extracted from differing baseline ranges (angular scales). The best-fit model is overlaid in red.
\label{fig:lores-3}}
\end{figure}

\begin{figure}
\plotone{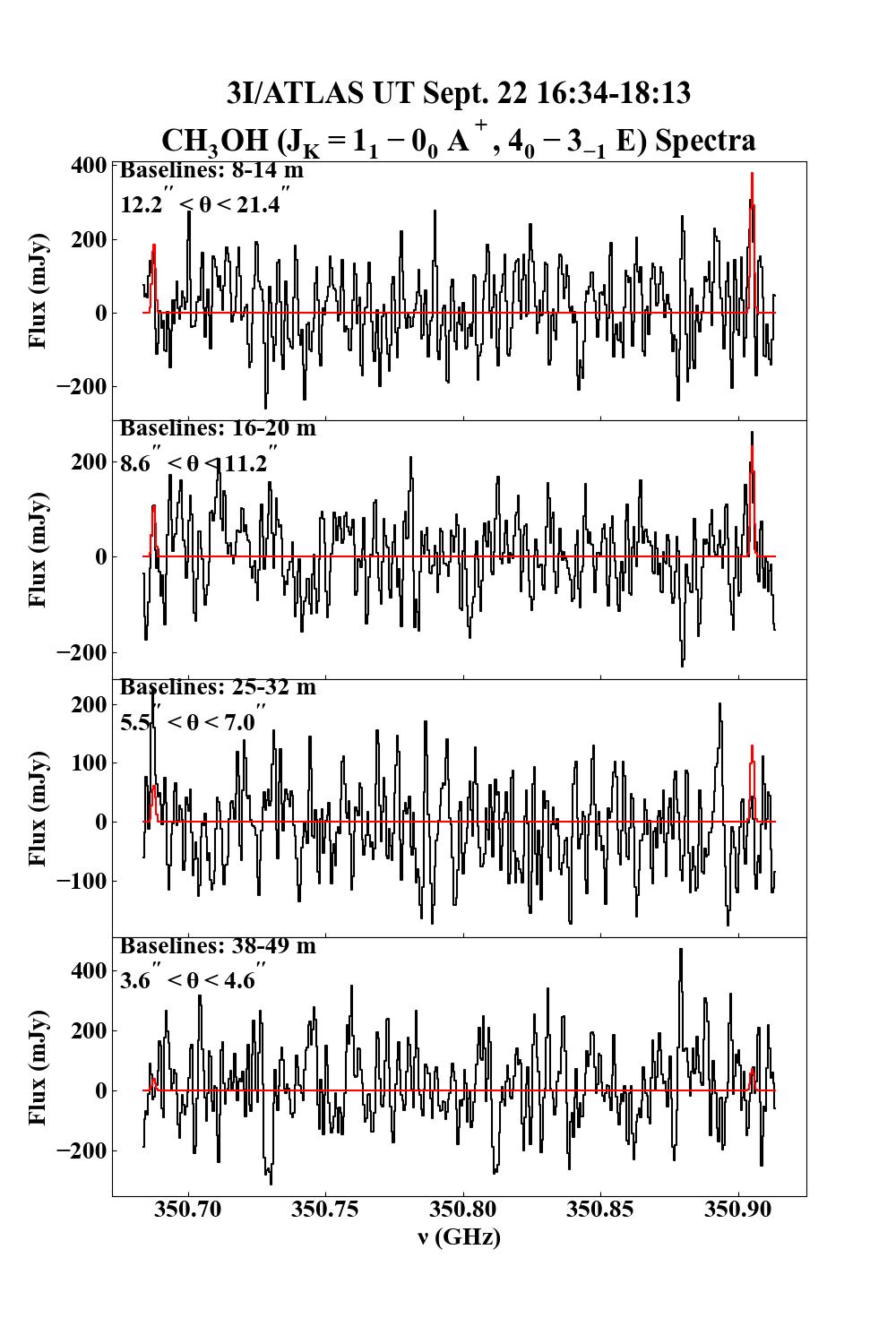}
\caption{CH$_3$OH spectrum in 3I on September 22, with each panel representing spectra extracted from differing baseline ranges (angular scales). The best-fit model is overlaid in red.
\label{fig:lores-4}}
\end{figure}

\begin{figure}
\plotone{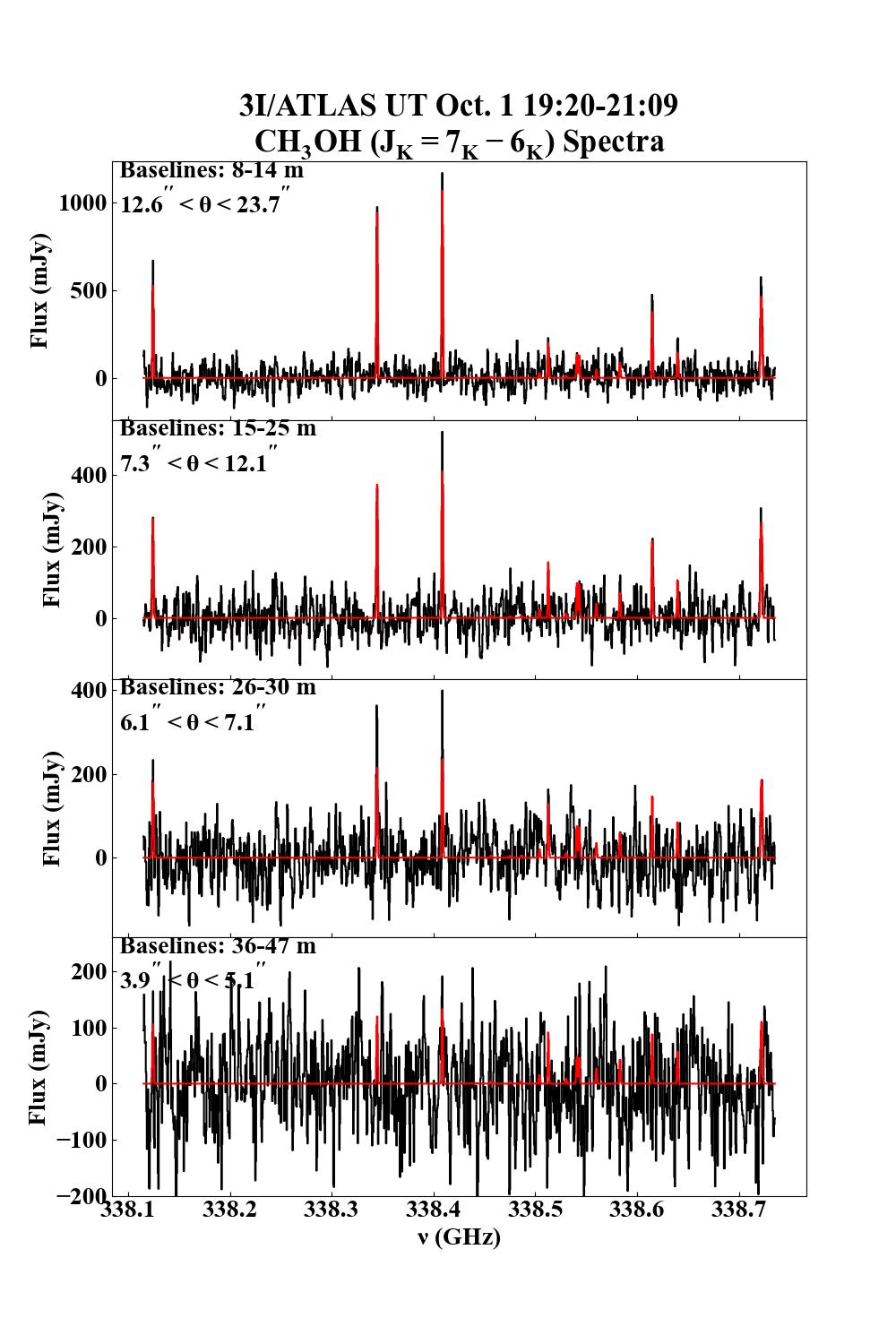}
\caption{CH$_3$OH spectrum in 3I on October 1, with each panel representing spectra extracted from differing baseline ranges (angular scales). The best-fit model is overlaid in red.
\label{fig:lores-5}}
\end{figure}

\begin{figure}
\plotone{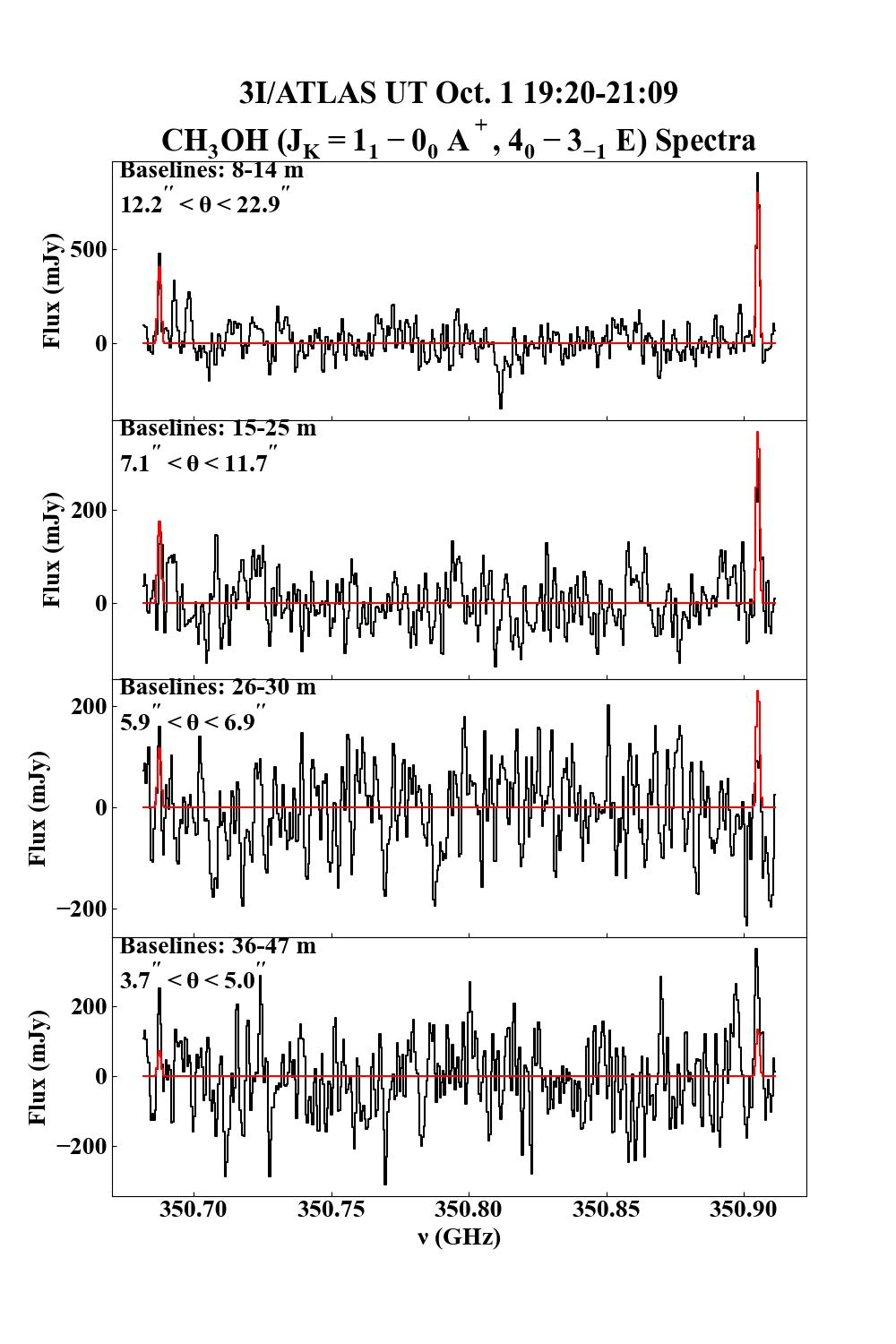}
\caption{CH$_3$OH spectrum in 3I on October 1, with each panel representing spectra extracted from differing baseline ranges (angular scales). The best-fit model is overlaid in red.
\label{fig:lores-6}}
\end{figure}

\begin{figure*}
\gridline{\fig{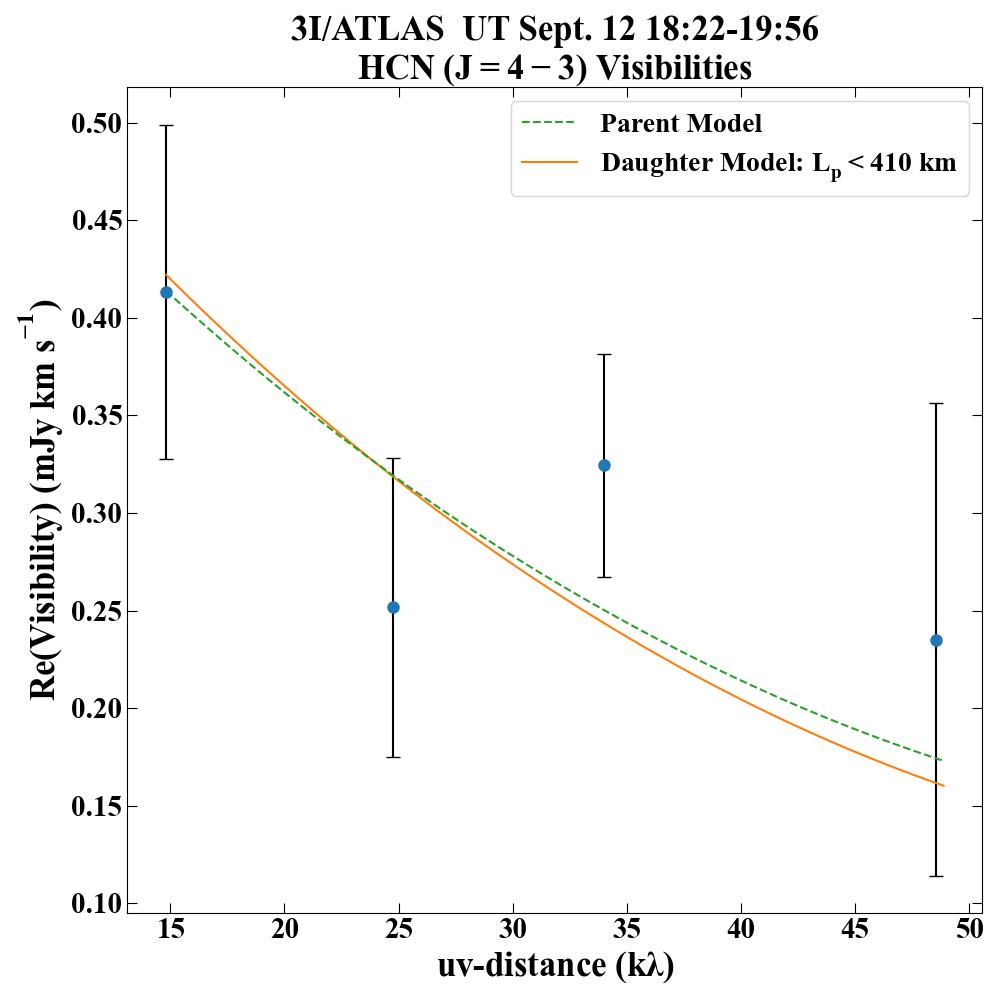}{0.45\textwidth}{(A)}
          \fig{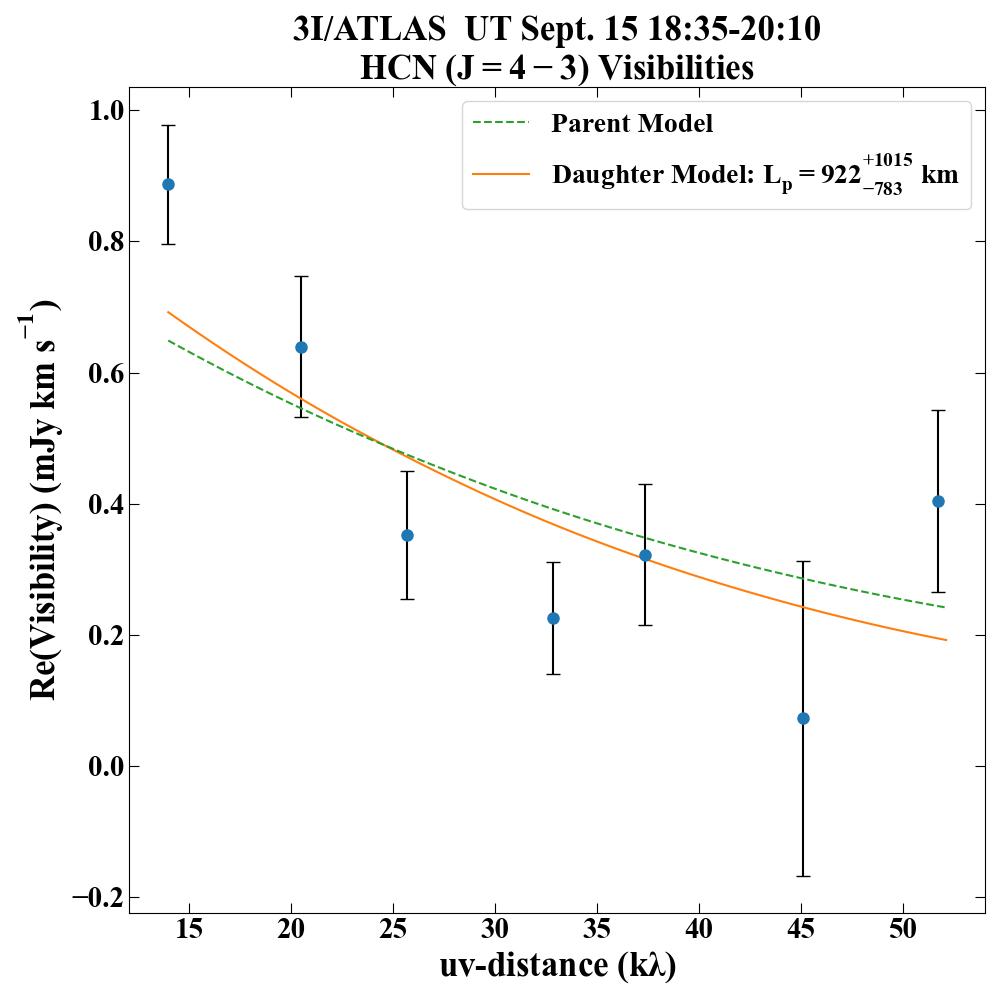}{0.45\textwidth}{(B)}
}

\caption{\textbf{(A)--(B).} HCN (\Ju{}=4--3) visibilities as a function of $uv$-distance (expressed in $k\lambda$) on September 12 and 15. The best-fit 3D daughter model and a parent models are shown.
\label{fig:hcn-vis}}
\end{figure*}

\begin{figure*}
\gridline{\fig{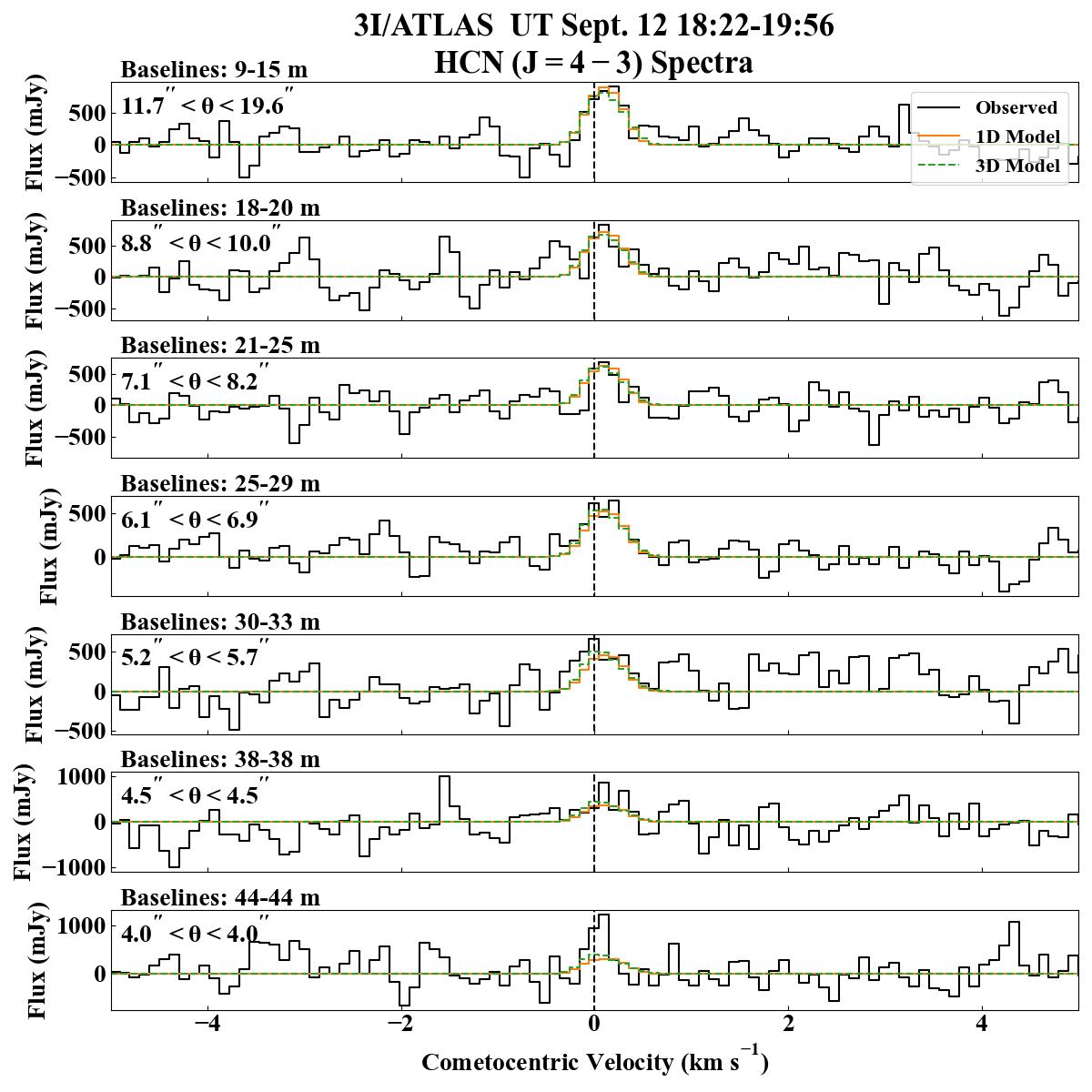}{0.45\textwidth}{(A)}
          \fig{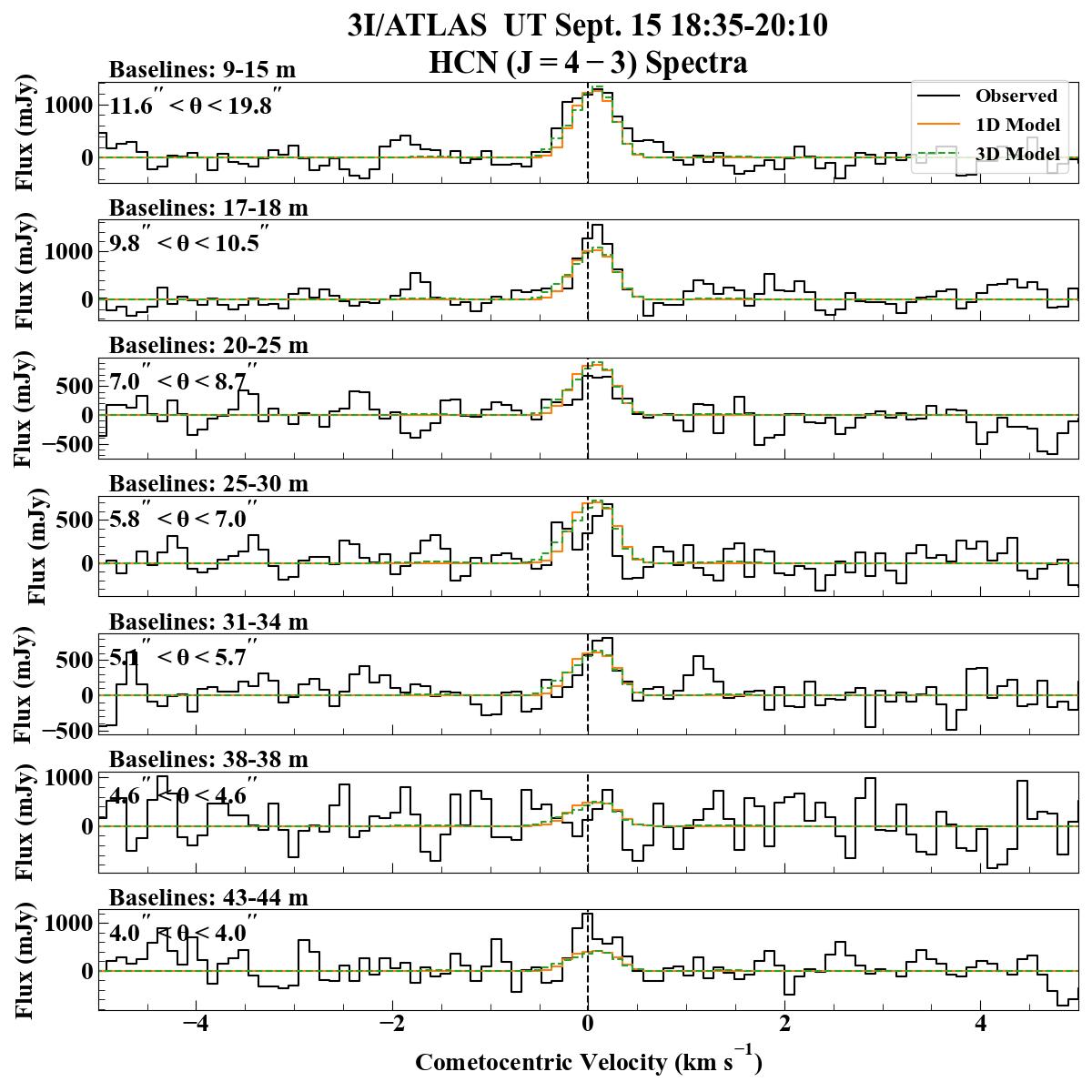}{0.45\textwidth}{(B)}
}
\caption{\textbf{(A).} HCN spectrum in 3I on September 12, with each panel representing spectra extracted from differing baseline ranges (angular scales). The best-fit 1D and 3D parent models are overlaid for comparison. \textbf{(B).} As in Panel A for September 15.
\label{fig:hcn-compare}}
\end{figure*}

\begin{figure}
\plotone{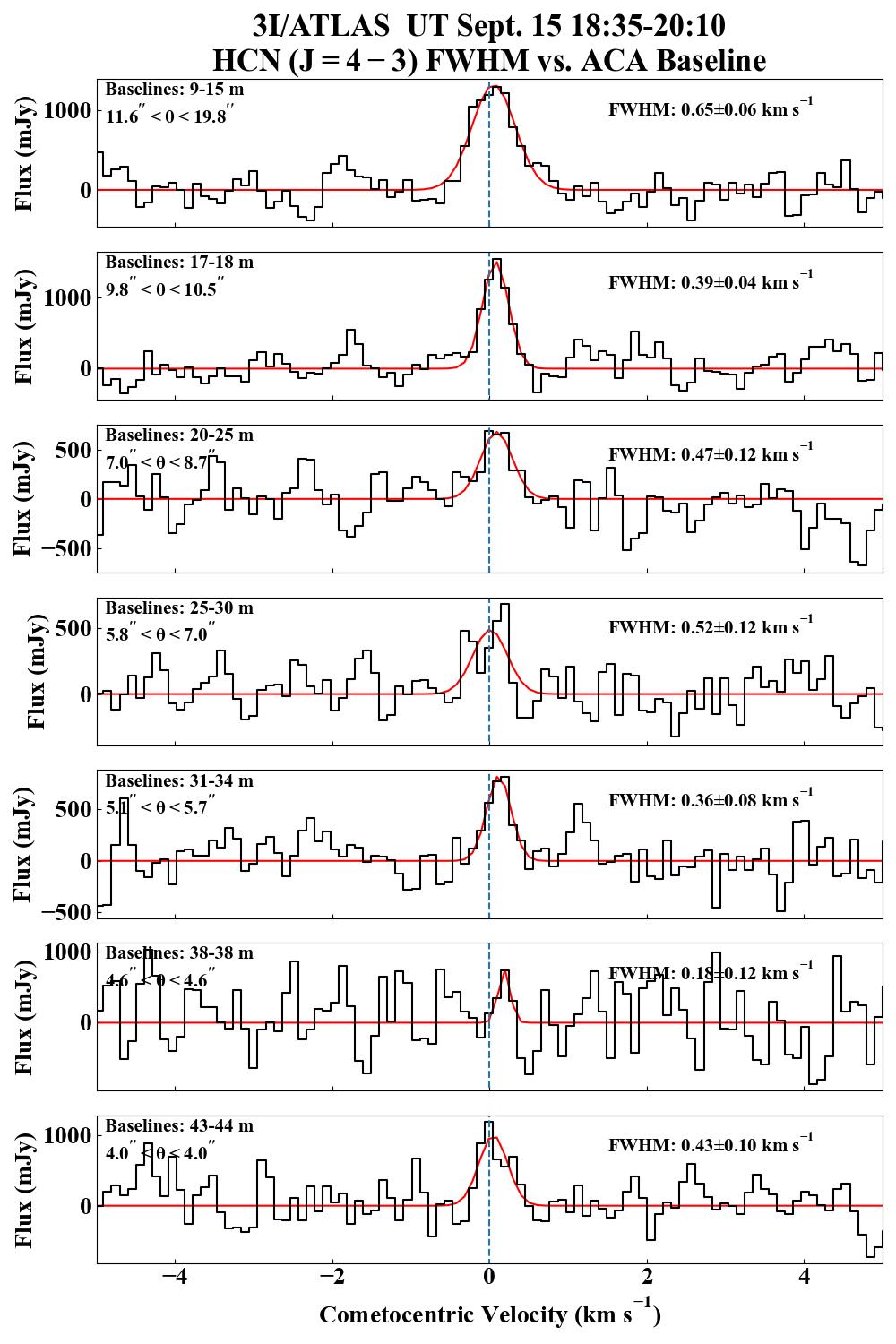}
\caption{HCN spectrum in 3I on September 15, with each panel representing spectra extracted from differing baseline ranges (angular scales). A best-fit Gaussian profile is overlaid in red, and the associated line FWHM is indicated.
\label{fig:hcn-fwhm}}
\end{figure}